\documentclass[11pt]{article}
\pdfoutput=1
\usepackage{soul}
\usepackage{jheppub}
\usepackage{amsfonts}
\usepackage{amsthm}
\allowdisplaybreaks[4]   
\usepackage{amsmath}
\usepackage{amssymb}   
\usepackage{placeins}
\usepackage{euscript}   
\usepackage{cleveref}    
\usepackage[dvipsnames]{xcolor}          
\usepackage{tensor}     
\usepackage{lipsum}  
\usepackage{graphicx}
\usepackage{caption}
\usepackage{afterpage}
\usepackage{subcaption}   
\usepackage{hyperref}
\usepackage{enumerate}
\hypersetup{
	colorlinks,
	linkcolor =NavyBlue,
	citecolor=MidnightBlue,
	urlcolor=NavyBlue,
}
\newcommand{\bea}{\begin{eqnarray}}
\newcommand{\eea}{\end{eqnarray}}
\newcommand{\ba}{\begin{eqnarray}}
\usepackage{braket}
\newcommand{\ea}{\end{eqnarray}}

\newcommand{\beq}{\begin{equation}}
\newcommand{\eeq}{\end{equation} }
\newcommand{\beqa}{\begin{eqnarray}}

\newcommand{\eeqa}{\end{eqnarray}}
\newcommand{\beqar}{\begin{eqnarray*}}
\newcommand{\eeqar}{\end{eqnarray*}}
\newcommand{\e}[1]{\text{e}^{#1}}

\newcommand{\nmax}{{n_\text{max}}}
\newcommand{\mext}{{m_\text{ext}}}

\newcommand{\be}{\begin{equation}}
\newcommand{\ee}{\end{equation}}
\newcommand{\diff}{\mathrm{d}}

\newcommand{\rp}{r_+}
\newcommand{\psip}{\psi_+}
\newcommand{\Ls}{L_\star}

\newcommand{\Z}{\mathcal{Z}}

\newcommand{\QT}{{\text{QT}}}

\newcommand{\iu}{\text{i}} 

\makeatletter
\newcommand{\dal}{\mathop{\mathpalette\dal@\relax}}
\newcommand{\dal@}[2]{%
  \begingroup
  \sbox\z@{$\m@th#1\square$}%
  \dimen0=\fontdimen8
    \ifx#1\displaystyle\textfont\else
    \ifx#1\textstyle\textfont\else
    \ifx#1\scriptstyle\scriptfont\else
    \scriptscriptfont\fi\fi\fi3
  \makebox[\wd\z@]{%
    \hbox to \ht\z@{%
      \vrule width \dimen0
      \kern-\dimen0
      \vbox to \ht\z@{
        \hrule height \dimen0 width \ht\z@
        \vss
        \hrule height 2\dimen0
      }%
      \kern-2.5\dimen0
      \vrule width 2.5\dimen0
    }%
  }%
  \endgroup
}
\makeatother

\definecolor{shadecolor}{rgb}{.25,.25,.25}

\usepackage{caption}
\usepackage{subcaption}

\usepackage{multirow}
\usepackage{tikz}
\usetikzlibrary{decorations.markings,decorations.pathmorphing}
\tikzstyle{singularity}=[carmine,line width=0.5mm,decorate, decoration={zigzag,amplitude=2,segment length=6.17}]

\definecolor{carmine}{rgb}{0.59, 0.0, 0.09}
\definecolor{egyptianblue}{rgb}{0.06, 0.2, 0.65}
\definecolor{frenchlilac}{rgb}{0.53, 0.38, 0.56}
\definecolor{darkspringgreen}{rgb}{0.09, 0.45, 0.27}
\definecolor{ochre}{rgb}{0.8, 0.47, 0.13}

\usepackage{tikz-3dplot}
\tikzstyle{particle1}=[lightseagreen,line width=0.5]
\tikzstyle{particle2}=[ochre,line width=0.5]
\definecolor{lightseagreen}{rgb}{0.13, 0.7, 0.67}
\definecolor{fandango}{rgb}{0.71, 0.2, 0.54}

\title{Holographic explorations of regular black holes in pure gravity 
}

\author[a]{Monserrat Aguayo,}
\author[a]{Leonardo Gajardo}
\author[b,c]{Nicolás Grandi,}
\author[a,d]{Javier Moreno,}
\author[a]{Julio Oliva,}
\author[e]{Martín Reyes}

\affiliation[a]{Departamento de Física, Universidad de Concepción,\\ Casilla, 160-C, Concepción, Chile \vspace{0.1cm}}
\affiliation[b]{Departamento de Física,
Universidad Nacional de La Plata,\\ Casilla de Correos 67, La Plata, Argentina \vspace{0.1cm}}
\affiliation[c]{Instituto de Física La Plata, Consejo Nacional de Investigaciones Científicas y Técnicas,\\ Diagonal 113 esquina 63, La Plata, Argentina \vspace{0.1cm}}
\affiliation[d]{Departament de Física Quàntica i Astrofísica, Institut de Ciències del Cosmos\\ Universitat de Barcelona, Martí i Franquès 1, E-08028 Barcelona, Spain
\vspace{0.1cm}}
\affiliation[e]{Departamento de Física, Facultad de Ciencias,
Universidad Nacional Autónoma de México,\\ A.P. 70-542, CDMX 04510, México
\vspace{0.1cm}}

\date{\today}
\abstract{It has been recently established in \textit{Phys.Lett.B 861 (2025) 139260} that asymptotically flat black holes exist in pure gravity theories in dimension greater than four. We extend this result to asymptotically five dimensional Anti-de Sitter spacetimes and perform bottom-up holographic explorations: Hawking-Page phase transitions, quasinormal modes of black holes providing decay times of excitations in the finite temperature conformal field theory and U-shaped string probes leading to the quark-antiquark potential in the dual field theory. The gravitational action contains a bared, negative cosmological constant and an Einstein-Hilbert term, supplemented by an infinite series of higher-curvature terms of the family known as Quasi-topological (QT) Lagrangians, leading to a lapse function in terms of infinite series which can be resumed to an analytic function for different coupling choices. The QT terms are defined as those satisfying a Birkhoff's theorem and possessing a holographic c-function, and exist at all orders in the curvature, for arbitrary dimension greater than four. Besides presenting the general formulae, we explore the physics
of three specific cases leading to Hayward, Dymnikova and Bardeen-like regular black holes. We also found the field redefinition that permits recasting the $\alpha'^3 W^4$ correction of Type IIB supergravity on $\mathcal{A}_5\times S^5$ as a series of QT terms. As anticipated in \textit{JHEP 11 (2019) 062}, due to the presence of a bared AdS radius $L$, the perturbative field redefinition renormalizes Newton's constant by an additive term of the form $\alpha'^3L^{-6}R$ and induces lower order terms of the form $\alpha'^3L^{-4}\mathcal{R}^2$ and $\alpha'^3L^{-2}\mathcal{R}^3$, both of which can be simultaneously written as QT terms, without spoiling the structure of the highest curvature term of the form $\alpha'^3\mathcal{R}^4$. Before finishing, beyond holography, we also provide an extended thermodynamic setup, that via variations of the dimensionful $\alpha$ and $L$, allow obtaining the Euler relation between finite thermodynamical quantities for the black hole of the theory.}

\begin{document} 
\maketitle
\flushbottom
%%%%%%%%%%%%%%%%%%%%%%%%%%%%%%%%%%%%%
%\setcounter{tocdepth}{2}
%the line above sets the depth of the table of contents. {2} means it will display section and subsections only.
%{\small
%\setlength\parskip{-0.5mm} 
%\tableofcontents
%}

%\newpage
%%%%%%%%%%%%%%%%%%%%%%%%%%%%%%

\section{Introduction}

Recently, a novel family of regular, asymptotically flat black holes were constructed in dimensions higher than four~\cite{Bueno:2024dgm}. Remarkably, these black holes are supported by pure gravity, considering General Relativity (GR) augmented by an infinite series of terms constructed out from Lagrangians belonging to the Quasi-topological (QT) family.\footnote{Black holes in General Relativity supported by non-linear electrodynamics were originally presented in the seminal work \cite{Ayon-Beato:1998hmi} As in the more recent work \cite{Cisterna:2020rkc}, in all these models regularity at the origin is achieved for fixed values of the charge, and therefore the smoothness of the $r=0$ center is not generic.}. The latter are singled out by requiring that for generic values of the couplings, a Birkhoff's theorem must be fulfilled, namely assuming spherical symmetry (planar or hyperbolic), implies the existence of an additional timelike Killing vector in the region of outer communication of black holes. In this work we extend such black holes to the asymptotically AdS case with arbitrary topology, and perform bottom-up holographic explorations including first-order phase transitions (confinement-deconfinement in the dual theory \cite{Witten:1998zw}) between large black holes and thermal AdS, we compute the relaxation time of excitations in the dual finite temperature CFT via the computation of quasinormal modes in the bulk \cite{Horowitz:1999jd}, and finally introduce U-shaped strings in the bulk attached to the boundary, that permit obtaining the quark-antiquark potential in the field theory side \cite{Maldacena:1998im}.

The first QT term, cubic in the curvature, was discovered in~\cite{Oliva:2010eb}, where the role of Birkhoff's theorem played a central role. It was also presented in~\cite{Myers:2010ru} in the context of constructing higher-curvature terms in five dimensions that could support a richer holographic description of four-dimensional conformal field theories (CFTs). A quartic QT Lagrangian, involving four powers of the Riemann tensor in five dimensions, was identified in~\cite{Dehghani:2011vu}, while a quintic extension was later constructed in~\cite{Cisterna:2017umf}. In~\cite{Bueno:2019ycr}, it was shown that, starting at sixth order in curvature, higher-order QT densities can be expressed in terms of lower-order ones, and a covariant expression for the corresponding Lagrangian was provided for arbitrary dimensions higher than four, and in a more explicit manner in \cite{Moreno:2023rfl} in terms of a minimal set of building blocks---see also, e.g., \cite{Ahmed:2017jod,Bueno:2022res,Moreno:2023arp}.

QT gravity theories form a subset of the so-called Generalized Quasi-topological (GQT) gravities. GQT gravities are characterized by having second-order field equations for the metric in static and spherically symmetric spacetimes (SSS), and notably, they exist in four dimensions---with Einsteinian cubic gravity being the first example identified~\cite{Bueno:2016xff}. While the equations of motion for the QT subfamily are algebraic in the metric function, those corresponding to \textit{proper} GQT theory require numerical integration~\cite{Hennigar:2016gkm,Bueno:2016lrh}.  Remarkably the propagator of the
graviton on the (Anti-)de Sitter ((A)dS) solutions of GQT gravities reduces to that of GR with a renormalized Newton's constant. In particular, the algebraic nature of the equations in SSS makes QT gravities special, as the field equations, after a first integration, reduce to a polynomial equation whose degree matches the highest power of the Riemann tensor appearing in the Lagrangian. This structure closely resembles Wheeler's polynomial arising in Lovelock gravity \cite{lovelock1970divergence,Lovelock:1971yv,Wheeler:1985nh,Boulware:1985wk,Cai:2001dz,Padmanabhan:2013xyr}. Such polynomial structure for solution in Lovelock and QT gravities extends beyond spherical symmetry for NUTs and bolts as shown in \cite{Corral:2019leh} and \cite{Bueno:2018uoy}, as well as for other gravitational instantons \cite{Corral:2022udb,Corral:2025yvr}.

At the same time, GQT gravities have proven to be highly valuable in holography, as they provide a framework to probing CFTs whose correlators include the most general structure compatible with conformal invariance~\cite{Myers:2010jv,Bueno:2018xqc,Li:2018drw,Cano:2022ord}. This has enabled the discovery of new universal relations valid across arbitrary CFTs~\cite{Bueno:2018yzo,Bueno:2020odt,Bueno:2022jbl}, facilitated the study of holographic entanglement entropy~\cite{Dey:2016pei,Bueno:2020uxs,Caceres:2020jrf,Anastasiou:2022pzm}, and allowed for the investigation of generic phenomena in holographic transport and superconductivity~\cite{Mir:2019ecg,Mir:2019rik,Edelstein:2022xlb,Murcia:2023zok}. The success of this program has motivated studying extensions including non-minimally coupled matter fields, such as a $\mathrm{U}(1)$ gauge field~\cite{Cano:2020ezi,Cano:2020qhy,Bueno:2021krl,Cano:2022ord,Bueno:2022ewf,Bueno:2025jgc}. Many of these explorations were possible because of the simple structure of the static solutions of these theories.

It is well known that ultraviolet-complete (UV) frameworks for gravity, namely String Theory, predict higher-curvature corrections to GR. In this context, the Einstein-Hilbert Lagrangian appears as the leading term in an infinite series of corrections, with the string constant $\alpha'$ serving as a perturbative expansion parameter. A top-down approach to obtain such expansion involves the computation of stringy corrections to the Virasoro–Shapiro amplitude, which in turn constrains the structure of the effective Lagrangian reproducing the corresponding amplitude, order by order in $\alpha'$---see the seminal works \cite{Gross:1986iv,Gross:1986mw,Metsaev:1987zx} and references thereof. Despite the large amount of work on the topic, it is fair to stablish that the general structure of the $\alpha'$ corrections remains unknown. This perturbative framework inherently allows for field redefinitions, which, while leaving the S-matrix invariant, can alter the explicit form of higher-curvature corrections at a given order in $\alpha'$. Notably, \cite{Bueno:2019ltp} demonstrated that the purely gravitational sector of Type IIB supergravity (SUGRA) compactified on $\mathcal{A}_5 \times S^5$, including the $\alpha'^3$ correction, can be mapped—via field redefinitions—to a GQT theory. In fact, it was proven there that any higher-curvature correction to the Einstein-Hilbert action, provided it is algebraic in the Riemann tensor, can be recast as a GQTG theory. Moreover, it was argued that such field redefinitions may enable mapping algebraic higher-curvature corrections to strictly QT theories. Additional evidence supporting this idea is provided in the Appendix of \cite{Bueno:2024dgm}.

One of the goals of this paper is to present an explicit field redefinition that maps the $\alpha'^3$ correction of Type IIB SUGRA on $\mathcal{A}_5 \times S^5$ to a QT theory. As we show in Section 6, the $\alpha'^3 \mathcal R^4$ correction to GR in five dimensions can be rewritten in terms of the quartic QT term. Due to the presence of a cosmological constant term in the five-dimensional action of the form $-12/L^2$, this recasting also induces quadratic and cubic curvature terms, scaled respectively by $\alpha'^3/L^4$ and $\alpha'^3/L^2$, both of which can likewise be expressed in terms of QT combinations. Newton's constant also receives a perturbative $\alpha'^3/L^6$ correction. Whether the remaining QT terms fully capture or not the perturbative corrections of String Theory at any order is beyond the scope of the present work, but interesting to mention that in \cite{Cisterna:2018tgx} it was shown that the simple dimensional reduction of each of QT terms from dimension five to dimension four, give rise to
models that are remarkably simple in homogeneous and isotropic cosmological
scenarios, leading to a first order Friedman equation for the scale factor identified as Cosmological QT theories
\cite{Arciniega:2018fxj,Arciniega:2018tnn,Arciniega:2019oxa,Edelstein:2020lgv,Edelstein:2020nhg}. This is precisely the situation when considering T-duality invariants in the dimensionally reduced cosmological setup
\cite{Hohm:2019jgu}, suggesting a potential origin of these higher-curvature terms
from higher dimensions in a UV-complete setup.

It is important to mention that the regular black holes found in \cite{Bueno:2024dgm} are a
robust prediction of the theory, and do not require any sort of fine tuning
of the coupling $\alpha_{n}$ (see \cite{Bueno:2024dgm}). They can be formed via gravitational collapse \cite{Bueno:2024zsx,Bueno:2024eig,Bueno:2025gjg} provided the energy content of the initial data is above a certain threshold. When the energy content of the spacetime is below such threshold, no black hole is formed, a phenomenology that is captured by the static black hole solutions that exist above a given critical mass. Below that mass, the spacetimes describe a continuous spectrum of regular solitons with positive energy.

The paper is organized as follows: In Section~\ref{sec:2} we present the action and its asymptotically AdS, regular, black hole solutions, including the spherical, hyperbolic and planar horizons. We present the key argument that allows for the existence of these solution, which is common to the asymptotically flat case presented in \cite{Bueno:2024dgm}, and show the explicit form of the metric functions of the three particular cases we will study in the next sections, namely the Hayward, Dymnikova and Bardeen-like black holes. In Section~\ref{sec:3}, we present the general formulae for the mass, entropy and temperature of the regular black holes. We then particularize these expressions for the three mentioned cases and study phase transitions at fixed temperature between regular black holes and thermal AdS, focusing on the effect of the higher-curvature terms on the Hawking-Page temperature. Section~\ref{sec:4} contains the analysis of the quasinormal (QN) modes of massless scalar probes on the regular black holes of the three-families under study. In Section~\ref{sec:5} we compute the effect of the higher-curvature terms on the holographic $q\bar{q}$ potential, which is interpreted as the correction on this observable at finite t'Hooft coupling. All the previous sections are bottom-up explorations on holographic properties of the regular black holes. In Section~\ref{sec:6}, we take a top-down approach and provide the explicit field redefinition that allows recasting the $\alpha'^3$ correction to Type-IIB SUGRA on $\mathcal{A}_5\times S^5$ as a series of QT terms containing up to order $\alpha'^3 \mathcal{R}^4$ terms. Finally, in Section~\ref{sec:7}, we move beyond holography and show that in the realm of Generalized Thermodynamics, the thermodynamic quantities of the regular black hole solutions of the theory fulfill a finite relation, namely an Euler relation. Section \ref{sec:8} contains conclusions and comments.

% \textbf{Note:} We have been in contact with a second group who have also been exploring the thermodynamics of regular black holes in AdS, but in the general case of arbitrary dimensions. Due to the overlap of the work \cite{Hennigar:2025ftm} with our Sections 2 and 3, we have agreed to upload the manuscripts to the \texttt{ArXiv} the same day.

\section{Regular AdS-black holes supported by pure gravity}\label{sec:2}

Let us consider GR in $D$ dimensions supplemented with a negative cosmological constant and a series of QT terms
\begin{equation}\label{action}
 I_\QT=\frac{1}{16\pi G}\int\diff^{D}x\sqrt{-g}\left(R-2\Lambda+\sum_{n}^\nmax\alpha_n\Z_n\right)\,,
\end{equation}
where $\Z_n$ is the QT Lagrangian\footnote{Since the off-shell expression of $\Z_n$ is not relevant for our purposes we do not include it here, but it can be found in, e.g.,~\cite{Bueno:2019ycr,Moreno:2023rfl}.} of order $n$ \cite{Oliva:2010eb,Myers:2010ru,Dehghani:2011vu,Ahmed:2017jod,Cisterna:2017umf,Bueno:2019ycr,Bueno:2022res,Moreno:2023rfl,Moreno:2023arp},  $n_{\max}\in\mathbb N$ denotes the maximum power of the curvature present in the QT family, $\Lambda=-(D-1)(D-2)/(2L^2)$ is the bared cosmological constant in terms of the bared AdS radius $L$ and $\alpha_n$ are dimensionful couplings of dimension $L^{2n-2}$. As usual, in higher-curvature gravity---see, e.g.,~\cite{Garraffo:2008hu,Bueno:2016ypa,Bueno:2022lhf}---the AdS radius of a maximally symmetric solution of the theory, is shifted from the bared value $L$, and hereafter, we denote such value as $L_\star=L_\star(L,\alpha_n)$.

As mentioned above, the defining property of GQT gravities is the reduction of the generically fourth-order equations of motion to a second-order system in static, spherically (or planar/hyperbolic) symmetric spacetimes \cite{Bueno:2016xff,Bueno:2017sui,Hennigar:2017ego,Ahmed:2017jod,Moreno:2023rfl, Moreno:2023arp}. These spacetimes are described by the metric
\begin{equation}
\diff s^{2} = -f(r)\, \diff t^{2} + \frac{\diff r^{2}}{f(r)} + \frac{r^{2}}{L^2} \diff\Sigma_{D-2,k}^{2} \, , \qquad 
\diff\Sigma_{D-2,k}^{2} =
\begin{cases}
L^2 \diff\Omega_{D-2}^{2}\,, & \text{for } k=1\,,\\
\diff\mathbf{x}_{D-2}^{2}\,, & \text{for } k=0\,,\\
L^2 \diff\mathbf{\Xi}_{D-2}^{2}\,, & \text{for } k=-1\,.
\end{cases}
\end{equation}
In this setup, the metric function $f(r)$ satisfies a second-order differential equation, with the topology of the horizon encoded in the constant $k = 1, 0, -1$, corresponding to spherical, planar, and hyperbolic geometries, respectively.

In dimensions $D \geq 5$, there exist $n-2$ proper, \textit{inequivalent} GQT gravities at each order $n$ in curvature,\footnote{We refer to GQT densities as \textit{inequivalent}---or belonging to distinct equivalence classes---whenever their equations of motion are not related by linear combinations~\cite{Bueno:2022res,Moreno:2023rfl,Moreno:2023arp}.} whose equations of motion explicitly involve derivatives of the metric function $f(r)$. In contrast, the QT subfamily---which contains only a single equivalence class at each order $n$, possesses a polynomial equation of motion for the single metric function, and for the theory in \eqref{action} reads
\begin{equation}
\sum_{n=0}^{n_{\max}}\alpha_{n}r^{D-2n-3}\left(  k-f\left(  r\right)  \right)
^{n}=m\, , \label{WP}%
\end{equation}
where $m$ is an integration constant.\footnote{The same equation \eqref{WP} is obtained
in Lovelock theories, but in such case $n_{\max}$ is bounded from above as
$n_{\max}\leq\left\lfloor (D-1)/2\right\rfloor $ since for higher powers
of the curvature, Lovelock terms become topological or identically vanish.}

The recent work on regular black holes \cite{Bueno:2024dgm,DiFilippo:2024mwm,Frolov:2024hhe,Bueno:2024eig,Bueno:2024zsx,Fernandes:2025fnz,Fernandes:2025eoc} stands from the observation that near
$r=0$, the polynomial equation implies that for generic couplings $\alpha_{n}$
the behavior of the function $f\left(  r\right)  $ is%
\begin{equation}
f\left(  r\right)  \propto\frac{1}{r^{\frac{D-2n_{\max}-1}{n_{\max}}}}+...\ .
\end{equation}
Remarkably, such behavior gives rise to a de Sitter core if $n_{\max
}\rightarrow\infty$, which is possible for finite dimension $D$. Including a cosmological term in the action, the generalization of the equation for the metric function $f(r)$ obtained in \cite{Bueno:2024dgm}, reads\footnote{Note that the factor of $D - 2n$ accounts for the fact that, in even dimensions, the term with $n = D/2$ does not contribute to the characteristic polynomial $h(\psi)$~\cite{Bueno:2024zsx}. This is analogous to the case of Lovelock gravity, where the $n = D/2$ term corresponds to the Euler density in that dimension and thus does not affect the equation of motion for $f(r)$.}
\begin{equation}\label{eq:hpsi}
h(\psi)=\frac{1}{L^2}+\psi+\sum_{n=2}^{\infty}\frac{D-2n}{D-2}\alpha_{n}\psi^{n}=\frac{m}{r^{D-1}}\,,\quad \text{with }%
\psi:=\frac{k-f\left(  r\right)  }{r^{2}}\, ,
\end{equation}
and $h(\psi)$ is a characteristic polynomial which encapsulates many properties of the QT gravity solutions--- see e.g.~\cite{Camanho:2011rj,Bueno:2016ypa,Bueno:2020odt, Bueno:2022res,Bueno:2022lhf}. 

The precise form of the resummation
obviously depends on the form of the couplings $\alpha_{n}$, which would be
fixed if the whole QT series may arise from the low energy,
effective description of a UV-complete theory.

If we restrict ourselves to odd dimensions, then the $D - 2n$ coefficient does not extract one term from the sum, and the expression of $h(\psi)$ allows for simple solutions when particular $\alpha_n$ are chosen. In the remainder of this paper, we will focus on three solutions, which are controlled by the following characteristic polynomials
\begin{alignat}{2}
&h_\text{H}\left(\psi\right) =\frac{1}{L^2}+\frac{\psi}{1-\alpha\psi}\,, \quad &&\text{if} \quad \alpha_n=\frac{D-2}{D-2n}\alpha^{n-1}\,,\label{eq:hHayward}\\
&h_\text{D}\left(\psi\right) =\frac{1}{L^2}-\frac{\log(1-\alpha\psi)}{\alpha}\,, \quad &&\text{if} \quad \alpha_n=\frac{D-2}{n(D-2n)}\alpha^{n-1}\,,\label{eq:hDymnikova}\\
&h_\text{B}\left(\psi\right) =\frac{1}{L^2}+\frac{\psi}{\sqrt{1-\alpha^2\psi^2}}\,, \quad &&\text{if} \quad \alpha_n=\frac{(1-(-1)^n)(D-2)\Gamma\left(\frac{n}{2}\right)}{2\sqrt{\pi}(D-2n)\Gamma\left(\frac{D+1}{2}\right)}\alpha^{n-1}\,.\label{eq:hBardeen}
\end{alignat}
 Their corresponding metric function $f(r)$ is obtained by \eqref{eq:hpsi}, namely
\begin{align}
&f_\text{H}(r)=k-\frac{r^{D+1}-L^2 m r^2}{\alpha  r^{D-1}-L^2 \left(r^{D-1}+\alpha  m\right)}\, ,\label{eq:Hayward}\\
&f_\text{D}(r)=k+\frac{r^2}{\alpha } \left(\e{\alpha  \left(L^{-2}-m r^{1-D}\right)}-1\right)\,,\label{eq:Dymnikova}\\
&f_\text{B}(r)=k+\frac{r^2 \left(r^{D-1}-L^2 m\right)}{\sqrt{L^4 r^{2D-2}+\alpha ^2 \left(r^{D-1}-L^2 m\right)^2}}\,.\label{eq:Bardeen}
\end{align}
It is interesting to notice that in the spherically symmetric case, for $m=0$ the spacetime reduces to global AdS, for positive values of $m<m_{\text{ext}}$ there is a continuous family of regular solitons, for $m=m_{\text{ext}}$ the solution describes a extremal black hole and for $m>m_{\text{ext}}$ the geometry correspond to black holes with inner and Cauchy horizons. More details are given below. Whenever an event horizon exists in the three special cases in which we will focus, we will refer to them as Hayward \cite{Hayward:2005gi}, Dymnikova \cite{Dymnikova:1992ux}, and Bardeen-like \cite{bardeen1968non,Bueno:2024dgm} black holes, respectively.

\section{Black holes thermodynamics of regular black holes in AdS}\label{sec:3}

In \cite{Bueno:2024dgm}, it was shown that the thermodynamics of regular, asymptotically flat black holes in QT gravity can be obtained in a closed form. To achieve this, the authors utilized general expressions to determine the Abbot-Deser-Misner (ADM) mass $M$, the Hawking temperature $T$, and the Wald entropy $S$ for QT black holes, as established in \cite{Bueno:2019ycr,Bueno:2022res}. The temperature can be obtained by requiring the absence of conical singularities in the Euclidean continuation of the black holes, which leads to a fixed value of the Euclidean time period. The entropy can be obtained via Wald formula \cite{Wald:1993nt,Iyer:1994ys} since the theory is invariant under diffeomorphisms. Finally, the mass can be obtained from the $r^{-(D-3)}$ term in the asymptotic expansion of the $g^{rr}$ component of the metric, since for generic values of the couplings this term is non-vanishing, and the spacetime asymptotic behavior coincides with that of the Schwarzschild-Tangherlini solution \cite{Tangherlini:1963bw}. The expression for the mass that can be read from this term in the asymptotic expansion leads to a mass that satisfies the first law. When a bared cosmological constant is included in the action, and a generic constant curvature horizon of curvature $k$ is considered, the expressions for the mass $M$, the temperature $T$ and the entropy $S$ are given by
\begin{align}
    M&=\frac{(D-2)\rp^{D-1}\Sigma_{{D-2},k}}{16\pi G}h(\psip)\,,\label{eq:thermoM}\\
    T&=\frac{1}{4\pi \rp}\left(\frac{(D-1)\rp^2h(\psip)}{h'(\psip)}-2k\right)\,,\label{eq:thermoT}\\ 
    S&=-\frac{(D-2)\Sigma_{{D-2},k}k^{D/2-1}}{8G}\int\frac{h'(\psip)}{\psip^{D/2}}\diff \psip\,,\label{eq:thermoS}
\end{align}
where the AdS scale enters through the characteristic polynomial $h(\psi)$ as given in \eqref{eq:hpsi} and $\Sigma_{D-2,k}$ is the volume of the transverse space with metric $\diff \Sigma_{D-2,k}^2$. This slight modification in $M$, $T$, and $S$ naturally preserves the validity of the first law, which remains satisfied as
\begin{equation}\label{eq:firstlaw}
    \diff M=T\diff S\,.
\end{equation}

In what follows, we compute explicitly the thermodynamic quantities for three classes of regular black holes, these are: the Hayward \cite{Hayward:2005gi}, Dymnikova \cite{Dymnikova:1992ux} and Bardeen-like \cite{bardeen1968non,Bueno:2024dgm} black holes for asymptotic AdS behavior and horizon topologies.

\subsection{Hayward black hole}

We start considering the metric function \eqref{eq:Hayward}. Depending on the value of $m$, the resulting metric represents the maximally symmetric vacuum solution, a continuous spectrum of solitons or a $D$-dimensional generalization of the Hayward-AdS black hole \cite{Hayward:2005gi}, which solves the theory \eqref{action} when $D$ is odd.

In the first case, when $m=0$, we recover a maximally symmetric solution,
\begin{equation}\label{eq:Hayasymp}
f_\text{H}\left(  r\right)\stackrel{m\rightarrow0}{=}k+\frac{r^2}{L^2-\alpha}=k+\frac{r^2}{L_\star^2}\, ,
\end{equation}
where the last equality identifies the effective AdS radius $L_\star=\sqrt{L^2-\alpha}$ of the solution \eqref{eq:Hayward} due to the presence of the higher-curvature terms in the action.  

From \eqref{eq:Hayasymp}, we conclude that the condition $L^{2} > \alpha$ is required for the black hole to exhibit AdS asymptotics, which aligns with the interpretation of $\alpha$ as a ultraviolet correction and $L^{2}$ as an infrared effect.

Of course, whenever the metric function \eqref{eq:Hayasymp} has at least one zero, it describes the geometry of a black hole. In this case, we denote the outermost root $r_+$ as the radius of the event horizon, and exploiting the relation $f(r_+)=0$, we can express the integration constant $m$ in terms of the horizon radius as
\begin{equation}
\frac{m}{\Ls^{D-3}}=\frac{y_+^D(1-\beta)^{\frac{D-3}{2}}(y_+^2+k(1-\beta))}{y_+^2-k\beta}\,,
\end{equation}
where we defined $\beta = \alpha/L^2=\alpha/(\alpha+\Ls^2)$ and $y_+ = r_+/L=r_+\sqrt{1-\beta}/\Ls$. We will keep these dimensionless parameters for the rest of the discussion, as they are more convenient.

When the integration constant $m$ is larger than zero but smaller than a critical value $m_\text{ext}$, given in units of the AdS radius by\footnote{The expression for $\mext$ is obtained by determining the radius of the extremal black hole, i.e., solving for $y_+$ while imposing a vanishing temperature, using the expression for the temperature given in \eqref{eq:Haywardtemp}. This yields
\begin{equation}\label{eq:Haywardrext}
    y_{+,\text{ext}}=\frac{\sqrt{k \left(3-2 \beta +D(-1+2 \beta )+\sqrt{(D-3)^2+8 (D-1) \beta }\right)}}{\sqrt{2(D-1)}
   }\,, \quad\text{ such that } T=0\,.
\end{equation}}
\begin{align}
\frac{\mext}{\Ls^{D-3}}=&\frac{k^{\frac{D-1}{2}}}{2^{\frac{D+3}{2}}\beta(1-\beta)^{\frac{D-3}{2}}  (D-1)^{\frac{D-1}{2}}}\left(D-3+4 \beta +\sqrt{(D-3)^2+8 (D-1) \beta }\right)\notag\\
&\times\left(3-D+2
   (D-1) \beta +\sqrt{(D-3)^2+8 (D-1) \beta }\right)^{\frac{D-1}{2}}\,,
\end{align}
the metric function \eqref{eq:Hayward} describes a soliton. For now, we focus on the black hole, which exists provided $m\geq \mext$. For $m=\mext$, the solution represents an extremal black hole, while for larger values of $m$, the spacetime features both a Cauchy horizon and an event horizon of radius $y_+$. 

For masses exceeding this threshold, we can compute the thermodynamic properties of the Hayward black hole using the general expressions \eqref{eq:thermoM}-\eqref{eq:thermoS} and particularizing $h(\psi)$ as given in \eqref{eq:hHayward}. For previous studies on the thermodynamics of the Hayward black hole see \cite{Myung:2006qr,Tharanath:2014naa,Dymnikova:2018uyo}. 

The Temperature can be obtained from \eqref{eq:thermoT}, yielding
\begin{equation}\label{eq:Haywardtemp}
   \Ls T= \frac{\sqrt{1-\beta}\left(k(D-3)y_+^2+(D-1) \left(y_+^4-k\beta  \left(k\left(1-\beta\right) +2 y_+^2\right)\right)\right)}{4 \pi
 y_+^3}\, .
\end{equation}
Regarding the ADM mass, we find that it is given by
\begin{equation}
    \frac{M}{\Ls^{D-3}}=\frac{(D-2)\Sigma_{D-2,k} y_+^{D-1}}{16 \pi  G(1-\beta)^{\frac{D-3}{2}}}\frac{\left(k(\beta-1)-y_+^2\right) }{\left( k\beta-y_+^2\right)}\, .
\end{equation}
Finally, we compute the Wald entropy, which admits a closed form expression in terms of Gauss hypergeometric functions, given by
\begin{equation}\label{eq:SHayward}
    \frac{S}{\Ls^{D-2}}=\frac{\Sigma_{D-2,k} y_+^{D-2}}{4G(1-\beta)^{\frac{D-2}{2}}}{}_2F_1\left(2,1-\frac{D}{2},2-\frac{D}{2},\frac{k\beta}{y_+^2}\right)\,.
\end{equation}
With these ingredients, we verify the first law \eqref{eq:firstlaw} and construct the free energy of the solution, given by $F=M-TS$, namely
\begin{align}\label{eq:freeHayward}
    \frac{F}{\Ls^{D-3}}&=\frac{y_+^{D-5}\Sigma^k_{D-2}}{16 \pi G \left(k \beta -y_+^2\right)(1-\beta)^{\frac{D-3}{2}}} \Bigg((D-2) y_+^4 \left(k (\beta
   -1)-y_+^2\right)-\left(k \beta -y_+^2\right)\\
   &\times\left(k (D-3) y_+^2+(D-1)
   \left(y_+^4+k \beta  \left(k (\beta -1)-2 y_+^2\right)\right)\right) \,
   _2F_1\left(2,1-\frac{D}{2},2-\frac{D}{2},\frac{k \beta
   }{y_+^2}\right)\Bigg)\notag\,.
\end{align}
Remarkably, the Hayward black hole---unlike other cases---allows to find closed expressions for the thermodynamic quantities in terms of the horizon radius parameterized by $y_+$. 

\begin{figure}
\begin{center}
\includegraphics[width=1\textwidth]{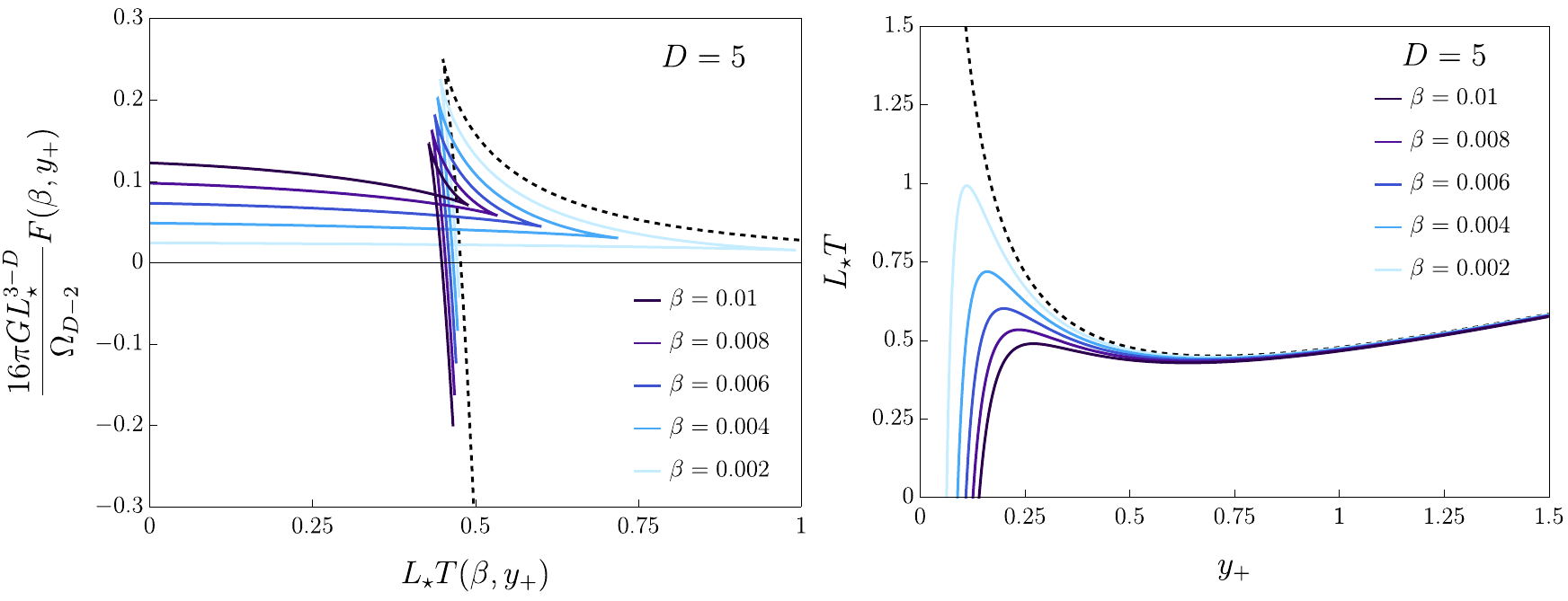}
\end{center}
\caption{(Left) We plot parametrically the free energy $F$ versus the temperature $T$ of the five-dimensional, spherically symmetric Hayward-AdS black hole for different values of the dimensionless coupling constant $\beta$ controlling the tower higher-curvature contributions. (Right) For the same black hole, we plot the Hawking temperature $T$ as a function of the dimensionless radius $y_+$ for different values of $\beta$. For non-vanishing $\beta$ there are three families of black holes, small ones connected to the extremal configuration at $T=0$, intermediate ones, and large ones. Only the former and the latter possess positive heat capacity, and when do not dominate, can be thought of as metastable configurations which decay via thermal fluctuation to thermal AdS. Above the Hawking-Page temperature, large black holes dominate.}
\label{fig:swallow5DHayward}
\end{figure}

To illustrate the different thermodynamic behavior of black holes in Einstein gravity and in the resummed tower of QT gravities, we plot in Figure~\ref{fig:swallow5DHayward} their respective free energy $F(\beta, y_+)$ as a function of the temperature $T(\beta, y_+)$ for several values of $\beta$. The dashed black curve corresponds to the Einstein case ($\beta = 0$), reproducing the standard AdS-Schwarzschild black hole, with a smooth and monotonic relation that features the well-known Hawking-Page transition between large black holes and thermal AdS, the latter possessing zero free energy at any temperature. In contrast, the colored solid curves represent the QT theories for increasing values of $\beta$, with a much richer structure. For finite $\beta$, multiple branches of solutions emerge, and sharp discontinuities appear in the free energy profile, indicating first-order phase transitions between different black hole phases. Notably, for sufficiently small $\beta$, the free energy becomes negative at lower temperatures than in the Einstein case, signaling an earlier onset of thermodynamic dominance for black hole configurations. 

At the same time, in Figure~\ref{fig:swallow5DHayward}, we show the temperature $ T $ of the $ D=5 $ black hole as a function of the dimensionless horizon radius $ y_+ $, for various values of $ \beta $. As soon as $ \beta $ is nonzero, up to three black holes can coexist at the same temperature. The range of temperatures for which this triple degeneracy occurs widens as $ \beta $ decreases. The radius of the extremal black hole in $ D=5 $ is obtained by evaluating expression~\eqref{eq:Haywardrext}, yielding $ \sqrt{4\beta + \sqrt{8\beta + 1} - 1}/2 $.

\subsection{Dymnikova-like black hole}
Let us now consider the metric function \eqref{eq:Dymnikova} obtained in the previous section with a convenient choice of $\alpha_n$. The AdS asymptotics of this spacetime can be obtained by taking $m=0$
\begin{equation}\label{eq:Dymnasymp}
f_\text{D}\left(  r\right)\stackrel{m\rightarrow0}{=}k+\frac{r^2}{\alpha}\left(\e{\alpha/L^2}-1\right)=1+\frac{r^2}{\Ls^2}\, ,
\end{equation}
where $\Ls^2=\alpha/(\e{\alpha/L^2}-1)$ is the effective AdS radius, which correct the bared radius $L$ due to the coupling $\alpha\geq 0$ of the higher-curvature terms. Unlike the Hayward-AdS case, there is no condition between $L$ and $\alpha$ for the solution to exhibit AdS asymptotics. However, in order to compare the thermodynamics of these cases, we will take the same values of $0<\beta=\alpha/L^2<1$ in all these spacetimes.

As we stated in the previous section, black hole configurations are given when the metric function \eqref{eq:Dymnikova} exhibits at least one zero, and for these cases, the outermost root $r_+$ represents the radius of the event horizon. As in the Hayward case, there is an extremal configuration in which both zeros of the function $f_{\text{D}}$ coincide. Expressing the integration constant $m$ in terms of the radius of the horizon yields
\begin{equation}
    \frac{m}{L_{\star}^{D-3}}=\left( \frac{e^{\beta}-1}{\beta} \right)^{\frac{D-3}{2}} \left[1-\frac{1}{\beta} \log{\left(1-k\frac{\beta}{y_+^2}\right)}\right]y_+^{D-1},
\end{equation}
where the same definitions of $y_+$ and $\beta$ have been used.

\begin{figure}[h!]
\begin{center}
\includegraphics[width=1\textwidth]{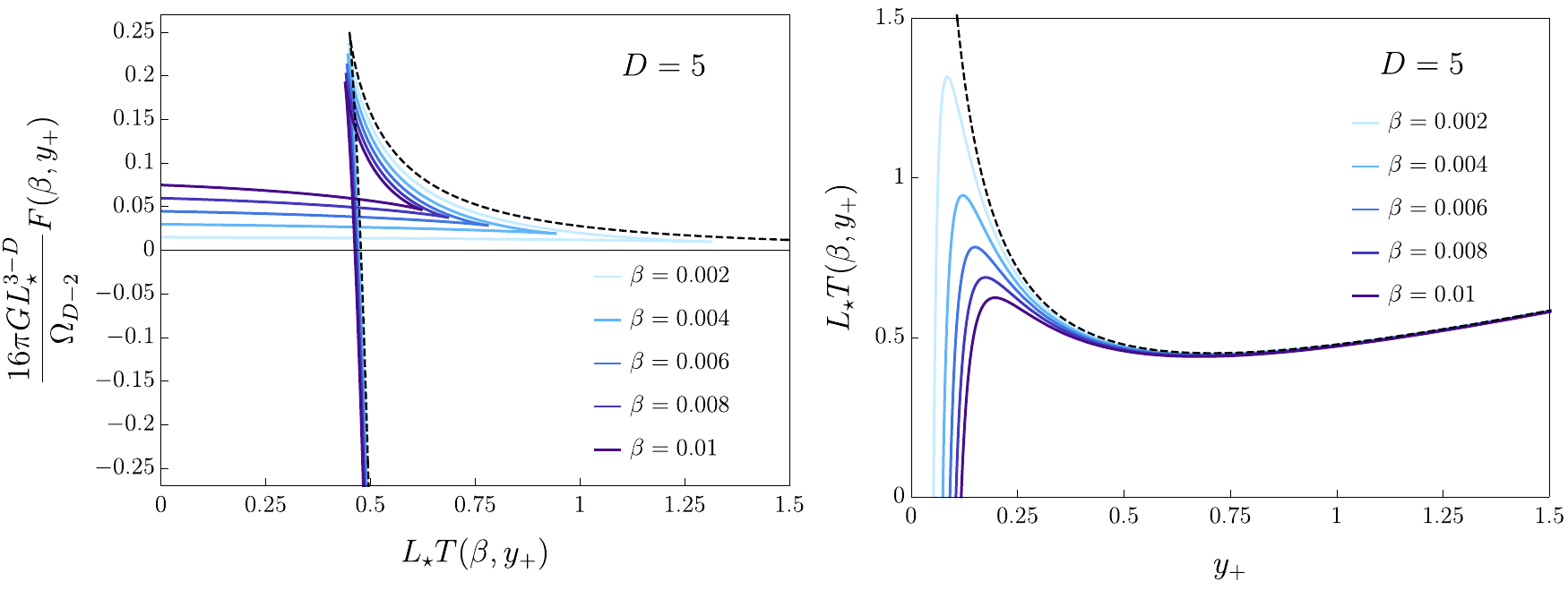}
\end{center}
\caption{(Left) We plot the free energy $F$ versus the temperature $T$ of the five-dimensional, spherically symmetric Dymnikova-AdS black hole for different values of the coupling $\beta$. (Right) The Hawking temperature $T$ vs the dimensionless radius $y_+$ for different values of $\beta$. As in the Hayward-like case, for non-vanishing $\beta$ there are three families of black holes, small ones connected to the extremal configuration at $T=0$, intermediate ones, and large ones. Only the former and the latter possess positive heat capacity. Above the Hawking-Page temperature, large black holes dominate. For a fixed value of the asymptotic AdS length, increasing the coupling $\beta$ decreases the critical temperature of the onset of the Hawking-Page phase transition between large black holes and thermal AdS.}
\label{fig:swallow5DDymnikova}
\end{figure}

The temperature can be computed using \eqref{eq:thermoT},  leading to
\begin{equation}
    L_{\star}T = \sqrt{\frac{\beta}{e^{\beta}-1}} \frac{(D-1)y_+^2+k[(D-1)\beta-2]+(D-1) \left( k-\frac{y_+^2}{\beta}\right)\log{\left( 1-k\frac{\beta}{y_+^2} \right)}}{4\pi y_+}\,.
\end{equation}
Unfortunately, in this case, one cannot obtain an analytical expression for the extremal value of $m$, because we have to deal with a transcendental equation to obtain the critical horizon radius by setting $T=0$:
\begin{equation}
    (D-1)y_{+,\text{ext}}^2+k[(D-1)\beta-2]+(D-1) \left( k-\frac{y_{+,\text{ext}}^2}{\beta}\right)\log{\left( 1-k\frac{\beta}{y_{+,\text{ext}}^2} \right)}=0\,,
\end{equation}
from which $m_{\text{ext}}$ is fixed. A not very illuminating expression for $m_\text{ext}$ can be given in terms of LambertW function. However, for our purposes, it is enough to find a numerical solution to this equation. The ADM mass is given by
\begin{align}
    \frac{M}{L_{\star}^{D-3}} &=\frac{(D-2)\Sigma_{D-2,k}y_+^{D-1}}{16\pi G} \left( \frac{e^{\beta}-1}{\beta} \right)^{\frac{D-3}{2}} \left[1-\frac{1}{\beta} \log{\left(1-k\frac{\beta}{y_+^2}\right)}\right] 
    \\&= \frac{(D-2)\Sigma_{D-2,k}}{16\pi G} m ,
\end{align}
which is proportional to the integration constant $m$, as usual. Finally, the Wald entropy results in
\begin{equation}
    \frac{S}{L_{\star}^{D-2}} = \frac{\Sigma_{D-2,k} y_+^{D-2}}{4G} \left( \frac{e^{\beta}-1}{\beta} \right)^{\frac{D-2}{2}} {}_2F_1\left( 1,1-\frac{D}{2},2-\frac{D}{2},\frac{k\beta}{y_+^2} \right)\ .
\end{equation}
With these expressions at hand one obtains the Helmholtz free energy of the solutions in terms of the horizon radius $y_+$, which is given by
\begin{align}
    \frac{F}{L_{\star}^{D-3}} =& \frac{\Sigma_{D-2,k}y_+^{D-1}}{16\pi G}\left( \frac{e^{\beta}-1}{\beta}\right)^{\frac{D-3}{2}}\left\{ (D-2)\left[ 1-\frac{1}{\beta}\log{\left(1-k\frac{\beta}{y_+^2}\right)} \right] \right.\nonumber\\
    &\left.- \left[(D-1)y_+^2+k[(D-1)\beta-2]+(D-1)\left( k-\frac{y_+^2}{\beta}\right)\ln{\left(1-k\frac{\beta}{y_+^2} \right)} \right] \right. \nonumber\\
    &\left. \times \ {}_2F_1\left( 1,1-\frac{D}{2},2-\frac{D}{2},\frac{k\beta}{y_+^2} \right) \right\}\ .
\end{align}
The plots for the free energy as a function of the temperature, for different values of the coupling $\alpha$ are given in Figure~\ref{fig:swallow5DDymnikova}. See the caption of such figure for details on the effect of the higher-curvature couplings on the onset of the Hawking-Page phase transition.
\subsection{Bardeen-like black hole}
\begin{figure}[h!]
\begin{center}
\includegraphics[width=1\textwidth]{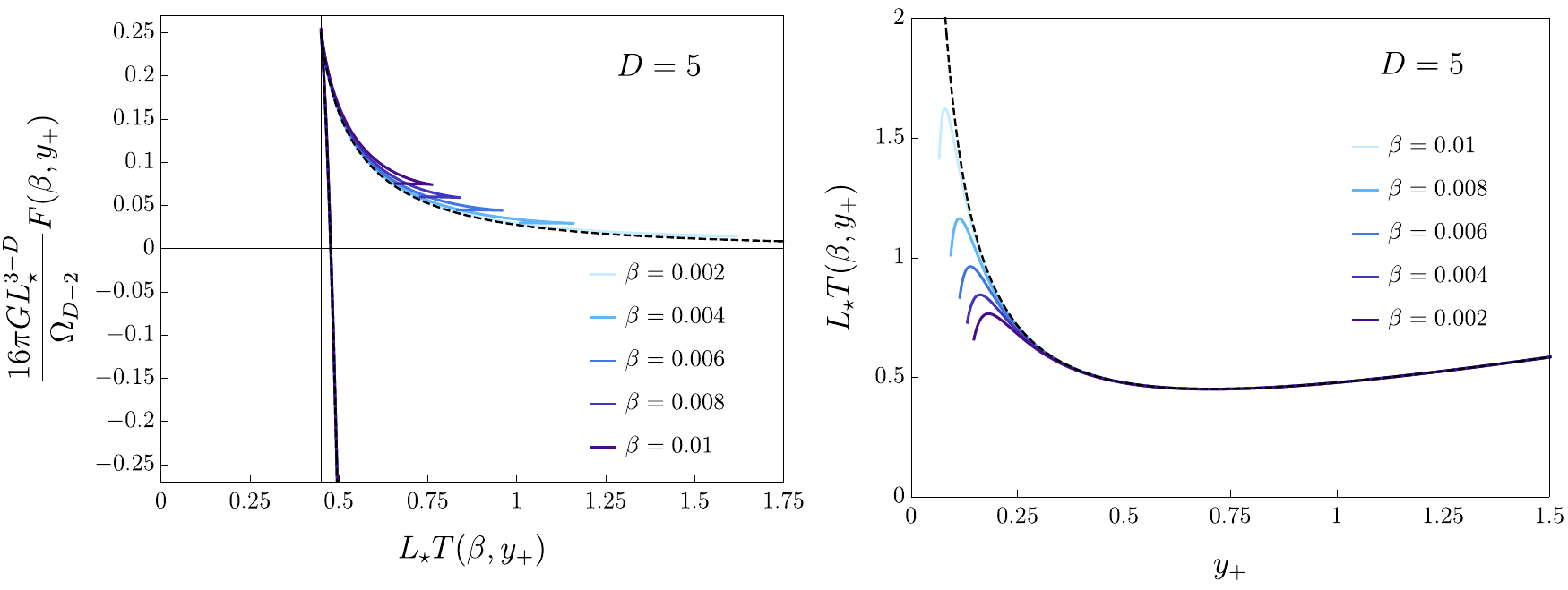}
\end{center}
\caption{(Left) We plot the free energy $F$ versus the temperature $T$ of the five-dimensional, spherically symmetric Bardeen-AdS black hole for different values of the coupling of the higher-curvature terms $\beta$. (Right) For the same black hole, we show the Hawking temperature $T$ as a function of the dimensionless radius $y_+$ for different values of $\beta$.}
\label{fig:swallow5DBardeen}
\end{figure}

The last black hole metric we are considering is the one constructed from the metric function \eqref{eq:Bardeen}, which corresponds to the choice of the couplings $\alpha_n$ given in \eqref{eq:hBardeen}. As we see below, the small black holes in this family behave completely different to their counterparts in the previous cases. The AdS asymptotic behavior in this case is
\begin{equation}\label{eq:Bardnasymp}
f_B\left(  r\right)\stackrel{m\rightarrow0}{=}1+\frac{r^2}{\sqrt{L^4+\alpha^2}}=1+\frac{r^2}{\Ls^2}\, ,
\end{equation}
where the effective AdS radius in this case is $\Ls^2=\sqrt{L^4+\alpha^2}$. The integration constant $m$ in terms of the horizon radius is given by
\begin{equation}
    \frac{m}{\Ls^{D-3}}=\frac{y_+^{D-1}}{(1+\beta)^{\frac{D-3}{4}}} \left(1+ \frac{k}{\sqrt{y_+^4-k^2\beta^2}} \right).
\end{equation}
Just as in the Dymnikova case, it is not possible to obtain a close expression for the extremal integration constant $m$ nor for the extremal radius. But, computing the temperature, which gives
\begin{eqnarray}
    \Ls T =(1+\beta^2)^{1/4}\frac{(D-1)\left( \left( y_+^4-k^2\beta^2 \right)^{3/2} -k^3\beta^2\right)+(D-3)ky_+^4}{4\pi y_+^5},
\end{eqnarray}
we can again obtain a equation for the extremal radius $y_{+,\text{ext}}$ by setting $T=0$
\begin{eqnarray}
    (D-1)\left( \left( y_+^4-k^2\beta^2 \right)^{3/2} -k^3\beta^2\right)+(D-3)ky_+^4=0\ .
\end{eqnarray}
The ADM mass for the Bardeen-like black hole is
\begin{eqnarray}
    \frac{M}{\Ls^{D-3}} &=& \frac{(D-2)\Sigma_{D-2,k}y_+^{D-1}}{16\pi G}\frac{1}{(1+\beta)^{\frac{D-3}{4}}} \left(1+ \frac{k}{\sqrt{y_+^4-k^2\beta^2}} \right) \\
    &=& \frac{(D-2)\Sigma_{D-2,k}}{16\pi G} m,
\end{eqnarray}
which again is proportional to the integration constant $m$. On the other hand, the Wald entropy in this configuration results in
\begin{eqnarray}
    \frac{S}{\Ls^{D-2}}=\frac{\Sigma_{D-2,k}y_+^{D-2}}{4G}\frac{1}{(1+\beta)^{\frac{D-2}{4}}} {}_2F_1\left( \frac{3}{2},\frac{1}{2}-\frac{D}{4},\frac{3}{2}-\frac{D}{4},\frac{k^2\beta^2}{y_+^4} \right)\ .
\end{eqnarray}
Finally, computing the Helmholtz free energy leads to 
\begin{eqnarray}
    \frac{F}{\Ls^{D-3}}&=&\frac{\Sigma_{D-2,k}y_+^{D-7}}{16 \pi G} \frac{1}{(1+\beta^2)^{\frac{D-3}{4}}} \left\{ (D-2) y_+^6\left(1+ \frac{k}{\sqrt{y_+^4-k^2\beta^2}} \right) - \right. \nonumber\\ 
    & & \ \ \left[ (D-1)\left( \left( y_+^4-k^2\beta^2 \right)^{3/2} -k^3\beta^2\right)+(D-3)ky_+^4 \right]\times \nonumber\\
    & & \ \ \left. 
    {}_2F_1\left( \frac{3}{2},\frac{1}{2}-\frac{D}{4},\frac{3}{2}-\frac{D}{4},\frac{k^2\beta^2}{y_+^4} \right)
\right\}
\end{eqnarray}
This free energy versus the temperature is depicted in Figure~\ref{fig:swallow5DBardeen}.  As in the Hayward-like case, for non-vanishing $\beta$ there are three families of black holes, small ones connected to the extremal configuration at $T=0$, intermediate ones, and large ones. Only the former and the latter possess positive heat capacity, and when do not dominate, can be thought of as metastable configurations which decay via thermal fluctuation to thermal AdS. We truncated the line of small black holes because in this Bardeen-like case, smaller black holes posses negative entropy and are pathological. It can also be seen that the behavior of the Hawking-Page temperature as the higher-curvature coupling increases is different than in the previous Hayward and Dymnikova black holes. In this case the temperature of the first order phase transition between large black holes and thermal AdS increases as the higher-curvature coupling increases. As shown in section \ref{sec:5}, the U-shaped string probe also behaves differently in the Bardeen case than in the Hayward and Dymnikova families.

\section{(Quasi-)normal modes of regular black holes}\label{sec:4}

\begin{figure}[h!]
\begin{center}
\includegraphics[width=0.4\textwidth]{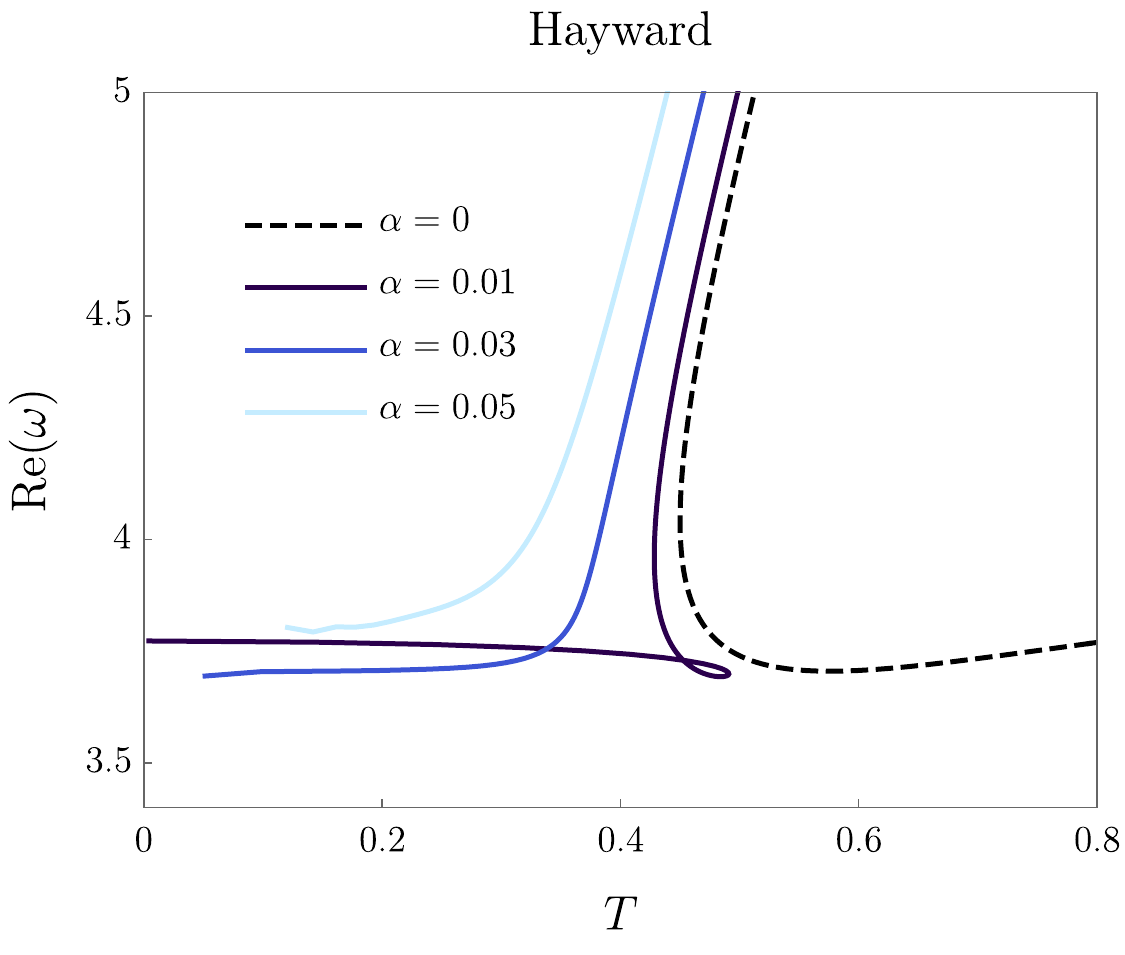}\qquad\includegraphics[width=0.4\textwidth]{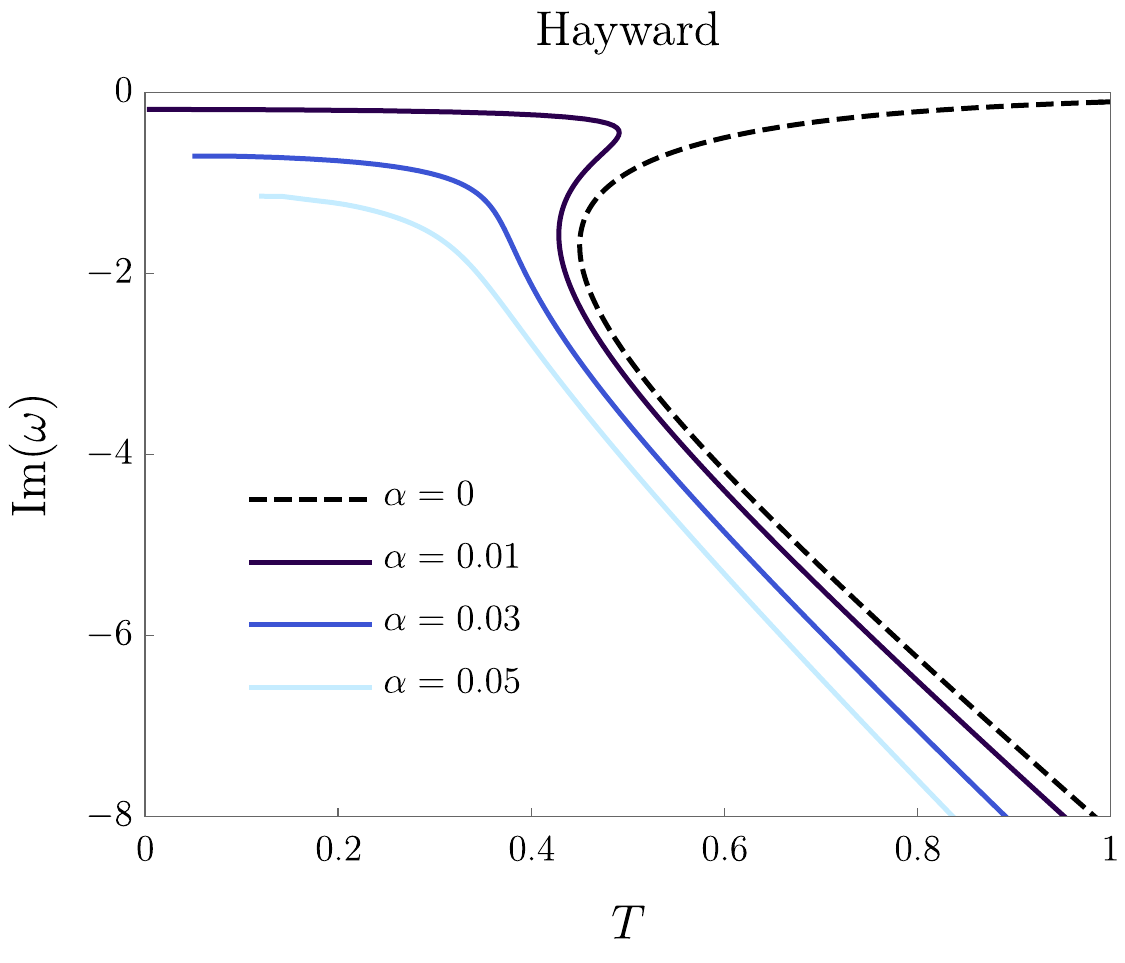}\\
\includegraphics[width=0.4\textwidth]{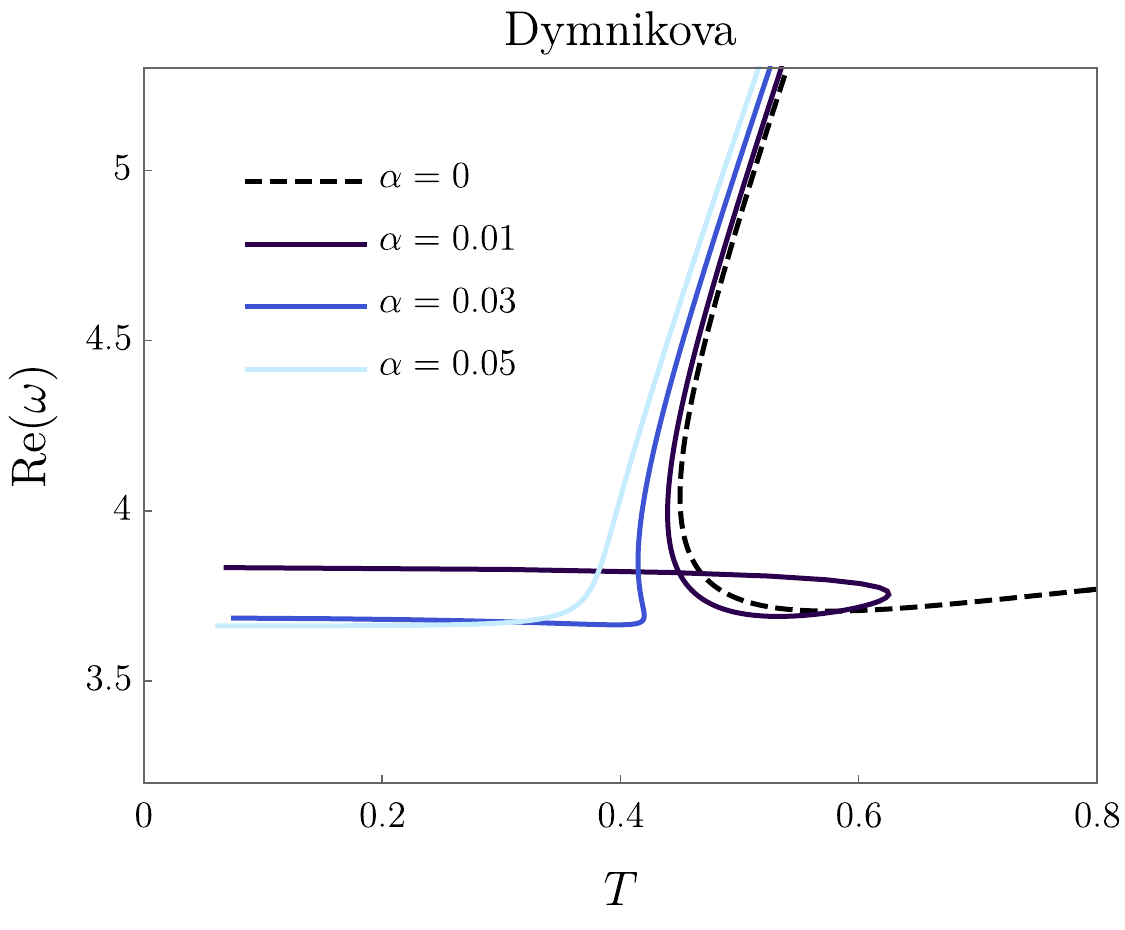}\qquad\includegraphics[width=0.4\textwidth]{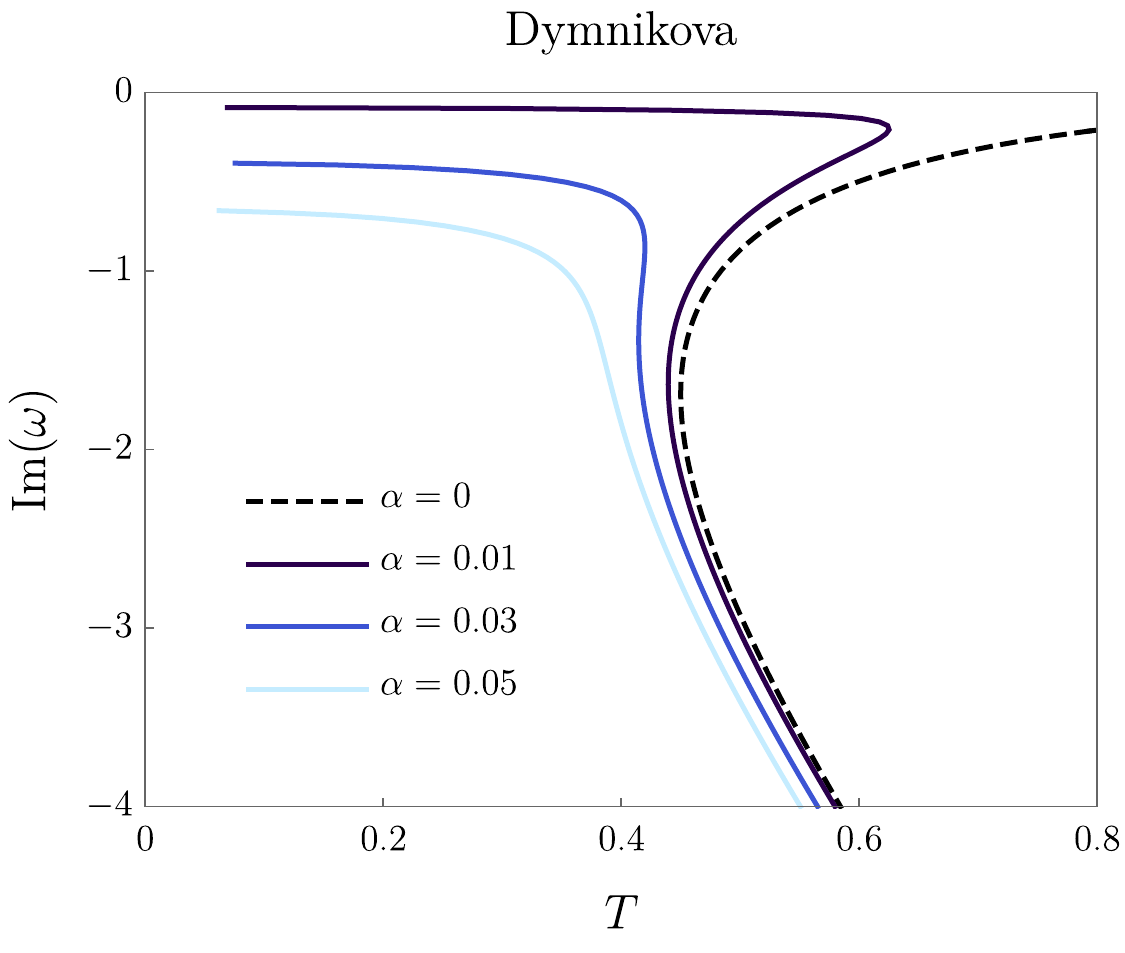}\\
\includegraphics[width=0.4\textwidth]{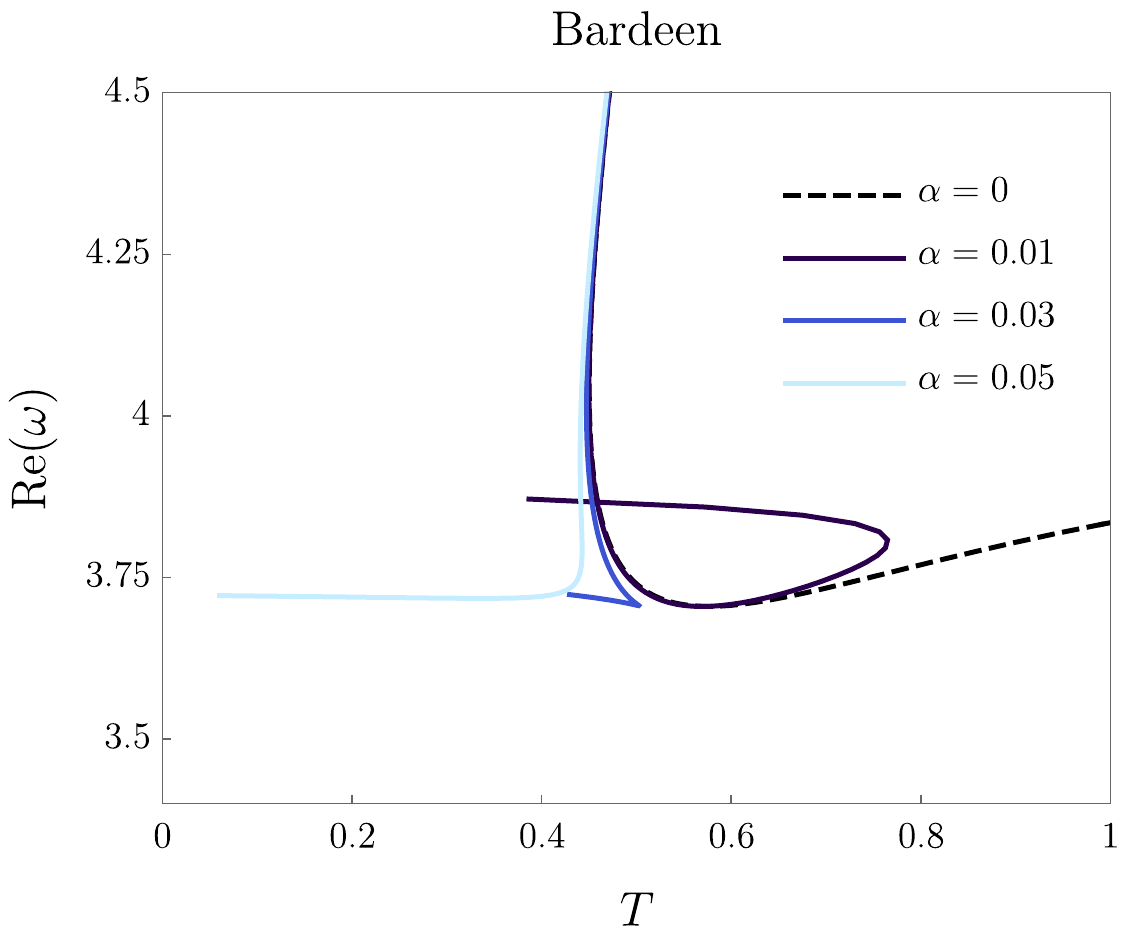}\qquad\includegraphics[width=0.4\textwidth]{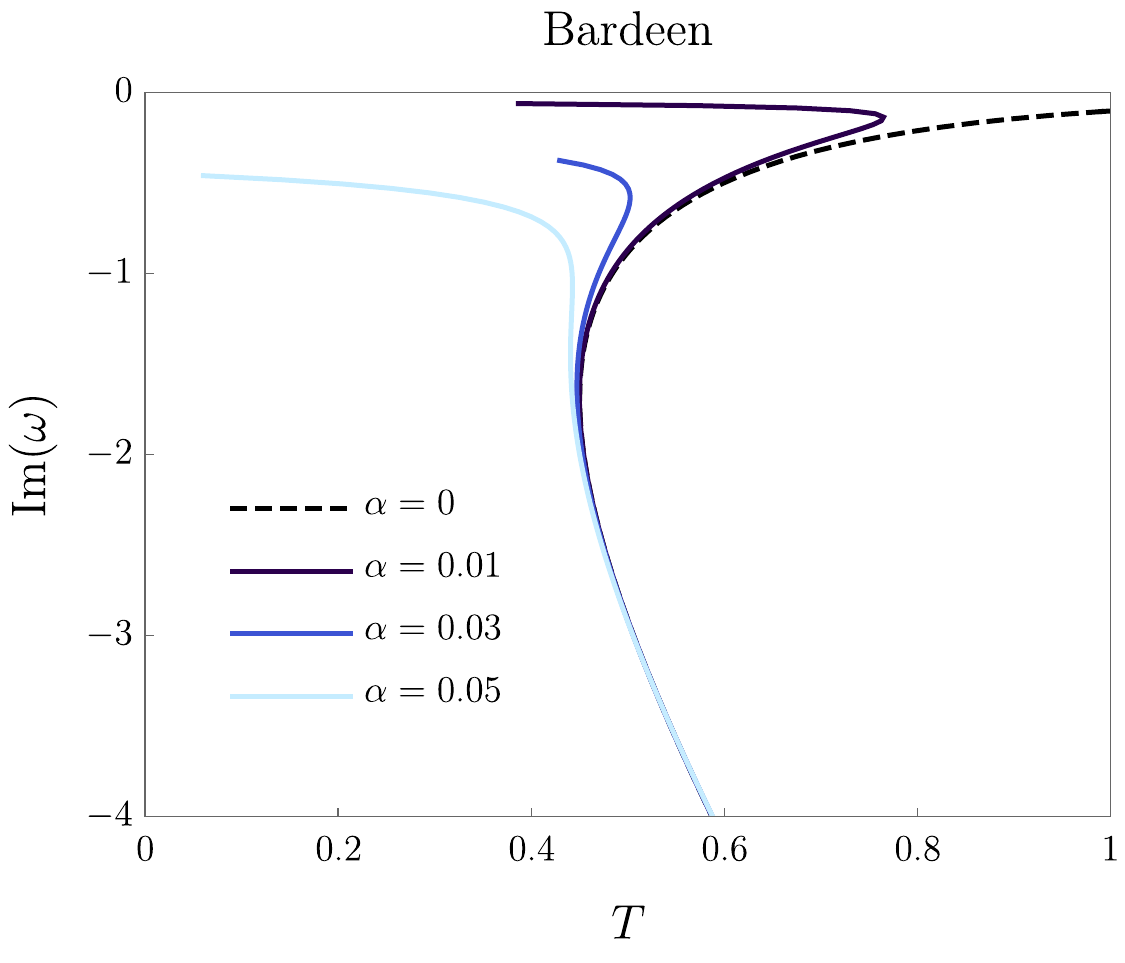}
\end{center}
\caption{\textbf{Regular black holes QNMs}. The real and imaginary part of the fundamental mode of the scalar probe, versus the temperature, varying $\alpha$. The dashed line corresponds to the Schwarzschild-AdS black hole, with a minimum, non-vanishing temperature, above which two black holes exist. For large black holes, the curves approach each other, since the effects of the corrections induced by $\alpha$ are suppressed. For small values of $\alpha$, as shown in the previous sections, there is a range of temperatures with three black holes. The curves should be continued for all values $T>0$, since the small black holes always start existing from an extremal configuration, not considered in our QNM analysis.}
\label{fig:QNMS}
\end{figure}

It is interesting to explore the asymptotically AdS black holes we have constructed in this work, with different types of probes. In this section, we study the behavior of massless scalar probes with ingoing boundary conditions at the horizon and Dirichlet boundary condition at infinity for the spherically symmetric, regular black holes in dimension five\footnote{The literature of QN modes of different types of pertubations of regular black holes is vast and a bibliographic summary is beyond the scope of this paper. However, the following works are particularly relevant for the present manuscript: references \cite{Koshelev:2024wfk,Koshelev:2024lyu} explored the QN modes of regular black holes in an infinite derivative expansion of four dimensional GR, finding evidence of unstable modes for near Schwazschild black holes. Also reference \cite{Konoplya:2024hfg} is relevant, where QN modes of Bardeen and Hayward-like black holes in asymptotically flat QT black holes were explored, followed by the analysis of the Dymnikova-like asymptotically flat case in \cite{Konoplya:2024kih}.}. The complex frequencies of the scalar modes in this case are known as QN frequencies, and their imaginary part captures the relaxation time of excitations in the dual field theory \cite{Horowitz:1999jd,Birmingham:2001pj}. In order to study the propagation of the massless scalar probe%
\begin{equation}
\dal\Phi=0\ ,
\end{equation}
on the regular black hole geometries, it is useful to move to coordinates that
are regular at the horizon, namely the ingoing Eddington-Finkelstein
coordinate $\diff v=\diff t+\diff r/f\left(  r\right) $ such that the metric reduces to%
\begin{equation}
\diff s^{2}=-f\left(r\right)  \diff v^{2}+2\diff v\diff r+r^{2}\diff\Omega^{2}_{D-2}\, .
\end{equation}
On this spacetime, the massless scalar probe can be separated as%
\begin{equation}
\Phi=\e{-\iu\omega v}R\left(  r\right)  Y\left(  \Omega_{D-2}\right)\, ,
\end{equation}
where $Y\left(  \Omega_{D-2}\right)  $ are eigenfunctions of the Laplace operator on
the $S^{D-2}$ sphere, namely $\nabla_{\Omega_{D-2}}Y=-l\left(  l+D-3\right)Y$ leading to the following equation for the
 radial dependence
\begin{equation}
fR^{\prime\prime}+\left(Dr^{-1}f-2\iu\omega-2r^{-1}f+f'\right)  R^{\prime}+\left(  -\iu\omega
r^{-1}D+2\iu\omega r^{-1}-l\left(  l+D-3\right)r^{-2}\right)  R(r)=0\ .
\end{equation}
Notice that this equation is linear in $\omega$, due to the use of the ingoing  null time $v$, instead of the usual Schwarzschild-like time coordinate $t$. We focus on non-extremal black holes for which the near horizon behavior of the function $R(r)$ is given by
\begin{equation}
R(r)=C_1\left(1+\mathcal{O}(r-r_+)\right)+C_2(r-r_+)^{2\iu\omega/4\pi T} \left(1+\mathcal{O}(r-r_+)\right)\ ,
\end{equation}
where $T$ is the black hole temperature $T=f'(r_+)/4\pi$. Requiring analiticity of the scalar field at the horizon $r=r_+$ implies that we must set $C_2=0$. In Schwarzschild-like coordinates this corresponds to imposing ingoing boundary conditions, namely, the near horizon behavior of the scalar field goes as $\Phi\sim\exp[-\iu\omega(t+(4\pi T)^{-1}\log(r-r_+))]$. In $D=5$, the asymptotic expansion of the scalar field reads
\begin{equation}
R(r)=\frac{D_1}{r^{\Delta_+}}\left(1+\mathcal{O}(r^{-1})\right)+\frac{D_2}{r^{\Delta_-}}\left(1+\mathcal{O}(r^{-1})\right)
\end{equation}
with $\Delta_+=4$ and $\Delta_-=0$. Requiring normalizability for the modes, implies that $D_2$ must vanish, and therefore the field vanishes at infinity as $\Phi\sim r^{-\Delta_+}\sim r^{-4}$, $\Delta_{+}$ being the scaling dimension of the operator $\mathcal{O}_{\Delta_+}(x)$, dual to the bulk field $\Phi$, in the boundary theory. It is possible to connect the analytic behavior in the horizon with the Dirichlet boundary conditions at infinity, for discrete sets of the complex frequencies in $\omega\in \mathbb C$. As for the Schwarzschild-AdS black hole \cite{Chan:1996yk,Horowitz:1999jd}, for the regular black hole of the type Hayward, Dymnikova and Bardeen, the QN frequencies have to be found numerically. We will focuss on the s-wave, namely we set the angular momentum of the scalar field to zero, $l=0$. Using standard methods, we find that the imaginary part of the frequencies are negative. The mode with the largest imaginary part is the mode that lives longer when a generic scalar field configuration is excited. In consequence, we explore the behavior of such mode in the asymptotically AdS$_5$, regular black holes. The characteristic relaxation time of this mode, $\tau\sim1/|\text{Im}(\omega)|$ has an interesting behavior and defines the relaxation time of perturbations on the strongly coupled, finite temperature, dual field theory \cite{Horowitz:1999jd,Birmingham:2001pj}. Along the lines of such reference, we explore the behavior of the longest-lived mode, as a function of the black hole temperature, which is the temperature of the dual field theory. For the different families of black holes and values of the coupling $\alpha$, as shown in the previous sections, there might be ranges at which for a given temperature more than one black hole exist. In order to allow for a consistent comparison between different values of the coupling $\alpha$, we have fixed to one the value of the effective AdS radius of the asymptotic region, namely $L_\star=1$. To explore departures from GR, we focus on small values of the coupling $\alpha$, for each of the three families of black holes. The frequencies are presented in Figure~\ref{fig:QNMS}. For a given temperature, such figure depicts the real and imaginary part of the fundamental mode, and there can be one, two or three black holes for a given temperature, depending on the value of $\alpha$. Large black holes are clearly identified since for those spacetimes, the effects of the higher-curvature terms are suppressed and the curves tend to match. On the other hand, small black holes are also easy to identify, since when $\alpha$ is non-vanishing the small black hole branch must continue up to the vertical axis $T=0$ as a consequence of the fact that small black holes have a minimum radius achieved by the extremal configuration (in this work we do not analyze the QN frequencies at the extremal cases). When there is a third branch in the middle, we refer to those black holes as intermediate black holes. This analysis is a consequence of comparing frequencies for a given, fixed temperature. Intermediate black holes have negative heat capacity, likewise small black holes when $\alpha=0$. It is interesting to notice a universal behavior that emerges in the imaginary part of the fundamental mode. For black holes with positive heat capacity, namely large black holes when $\alpha=0$ or large and small black holes when $\alpha\neq 0$, as the temperature increases the decay time decreases, namely the scalar fundamental excitation lives longer for smaller temperatures. There is of course a direct parallel with the resistivity of metals, which decreases as the temperature drops. On the other hand, intermediate black holes with negative heat capacity have an anomalous behavior in this context. Regarding $\text{Re}(\omega)$, large black holes have a universal behavior that allows drawing the following interpretation: any notion of regularized volume of the exterior region of the black hole should decrease as the black hole radius increases, and in consequence one expects a higher pitch for the fundamental frequency of oscillations for larger black holes. One actually observes a linear behavior for $\text{Re}(\omega)$ as a function of the temperature---see also \cite{Horowitz:1999jd}.
\newpage
\newpage

\section{String probes and $q\bar{q}$ potential}\label{sec:5}
\begin{figure}[h!]
\begin{center}
\includegraphics[width=0.4\textwidth]{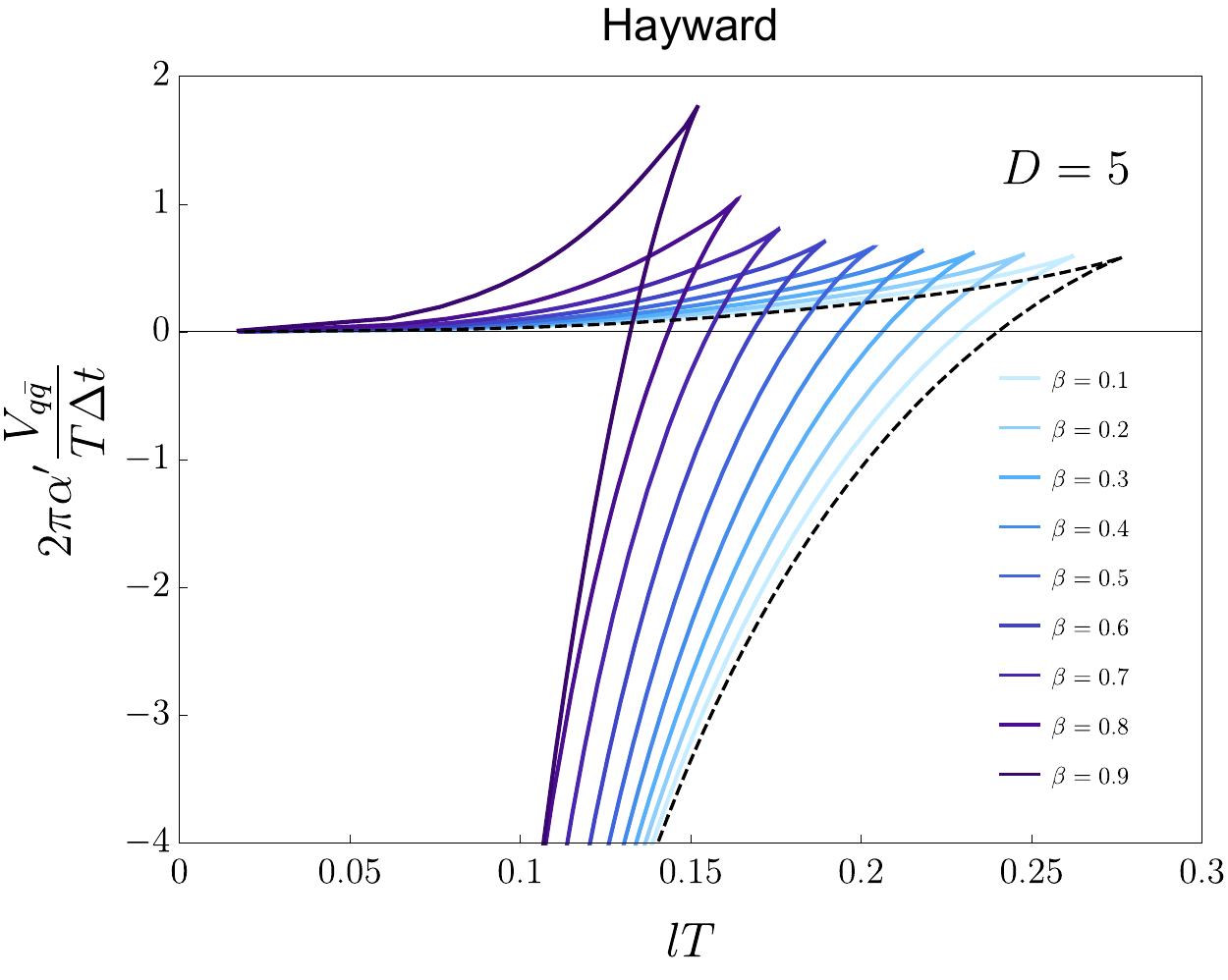}\qquad\includegraphics[width=0.4\textwidth]{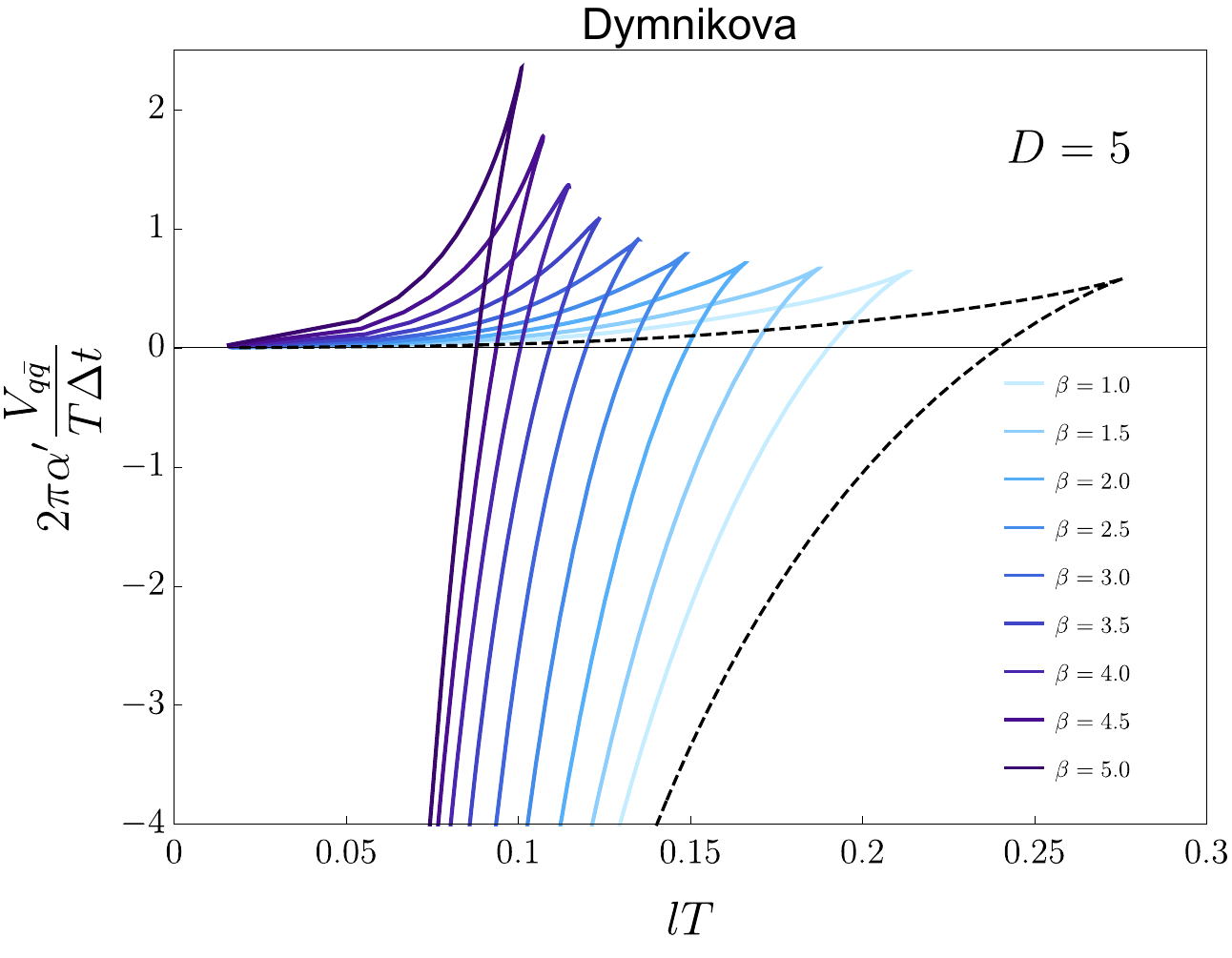}\\
\includegraphics[width=0.4\textwidth]{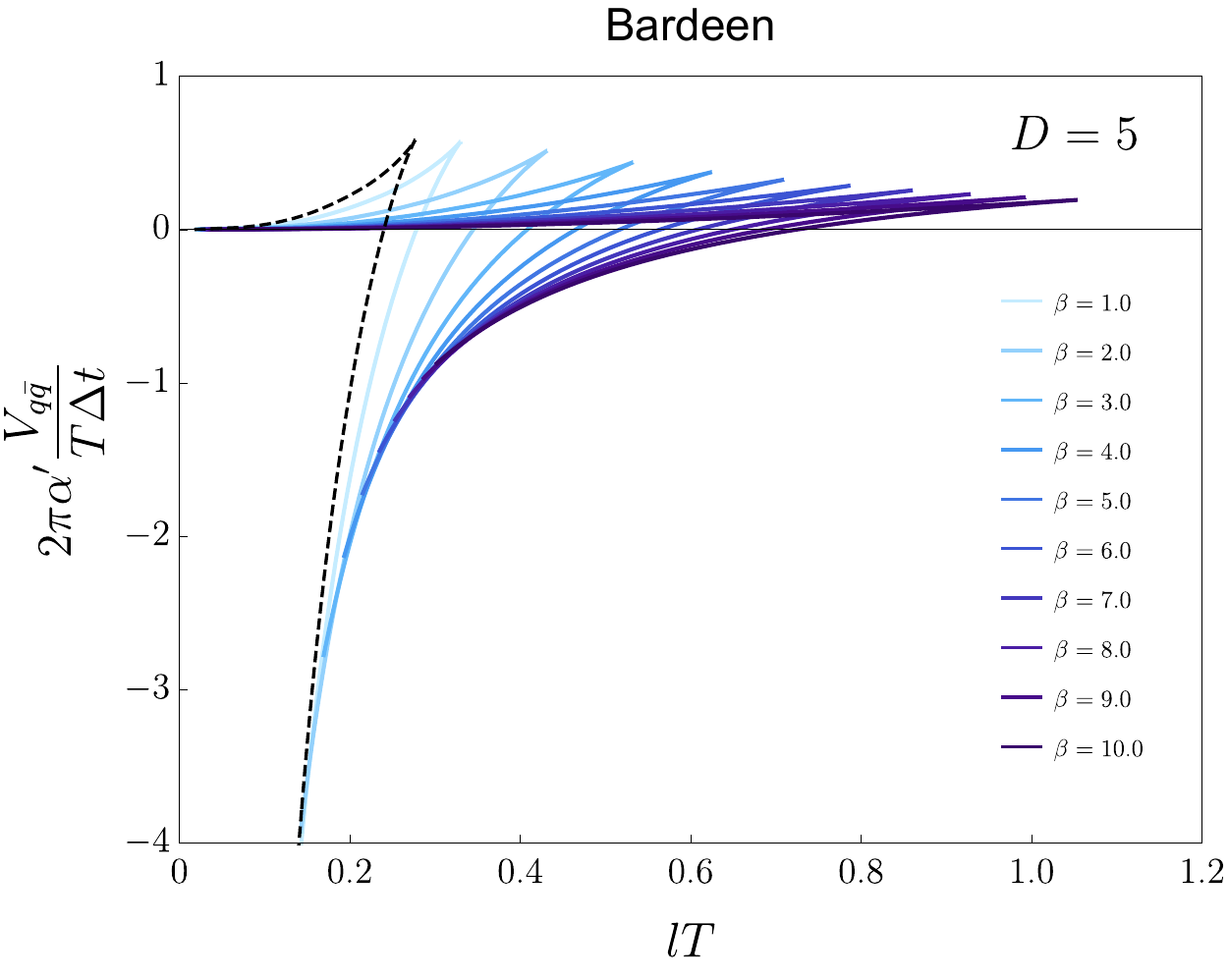}
\end{center}
\caption{\textbf{$V_{q\bar{q}}$.} Quark-antiquark potential at the boundary of planar, regular black holes, for different values of the coupling $\alpha=\beta L^2$. Both axis have been normalized with respect to the planar black hole temperature $T$ which is given by the universal relation $T=r_{+}/(\pi L^{2})$, regardless the presence of a non-vanishing $\alpha$.}
\label{fig:qq}
\end{figure}

It is known that string probes attached to the boundary capture the quark-antiquark potential, the quarks being described by the ends of the string, in the infinite mass limit. From field theory, the rectangular Wilson loop captures the $V_{q\bar{q}}$ potential via
\begin{equation}
\langle W(\mathcal{C})\rangle\sim e^{-\Delta tV_{q\bar{q}}}\ ,
\end{equation}
while from holography, the Wilson loop along the curve $\mathcal{C}$ can be obtained from a string whose ending points describe the trajectory $\mathcal{C}$, and explores the holographic radial direction. In the semiclassical limit, the Wilson loop can be obtained exponentiating the on-shell Nambu-Goto action \cite{Maldacena:1998im,Brandhuber:1998bs}, namely
\begin{equation}
\langle W(\mathcal{C})\rangle\sim e^{-I_{\text{NG}}}\ .
\end{equation}
Here, $\Delta t\rightarrow+\infty$. As with any geometrical quantity which, from the bulk, reaches the AdS infinity, one must regularize the infinities that emerge in the computation of the on-shell NG action. In the types of backgrounds we are considering, it is standard subtracting the NG action evaluated on two ``straight" string configurations exploring the bulk up to the horizon for black holes ($r_{\text{min}}=r_+$) or to the origin of the spherically symmetric solitons ($r_{\text{min}}=0$), respectively \cite{Maldacena:1998im}. Finally, the regularized on-shell action $I_{\text{reg}}(r_0)$ and the distance $l(r_0)$ of the $q\bar{q}$-pair for planar backgrounds or the angular separation for backgrounds with compact directions in the boundary, are written in terms of the value of the radial coordinate at which the string explores the IR the most, denoted by $r=r_0$ (for details we refer to \cite{Brandhuber:1998bs}). In order to be able to consider large separations between the $q\bar{q}$ pair, let us consider here the planar black hole background, namely

\begin{equation}
\diff s^2=-f(r)\diff t^2+\frac{\diff r^2}{f(r)}+r^2(\diff x^2+\diff y^2+\diff z^2),
\end{equation}
The on-shell action and the (angular) separation between the quarks is given by
\begin{equation}
\frac{I_{\text{reg}}(r_0)}{2\Delta t}=\int_{r_{0}}^\infty dr\left[\frac{f(r)^{1/2}r}{\sqrt{f(r)r^2-f(r_0)r_0^2}}-1\right]-\int_{r_\text{min}}^{r_0}dr\ ,
\end{equation}
and
\begin{equation}
l(r_0)=2\int_{r_0}^{\infty}\frac{r_0}{r}\left(\frac{f(r_0)}{f(r)}\right)^{1/2}\frac{dr}{\sqrt{f(r)r^2-f(r_0)r_0^2}}\ .
\end{equation}
These expressions lead to the results depicted in Figure~\ref{fig:qq}, for the quark-antiquark potential of planar, regular black holes of the Hayward, Dymnikova and Bardeen-like families.
Following the standard interpretation of the $q\bar{q}$ potential at finite temperature, we see that for small separation, namely $l$ small, there is a Coulomb-like behavior $V_{q\bar{q}}\sim-l^{-1}$ which exists regardless the value of the higher-curvature coupling $\alpha$. For small separations the turning point of the string approaches the boundary and the string does not notice the deformation of the geometry with respect to the planar AdS background. As the separation increases, there is a precise value of $l$, that depends on the value of $\alpha$, at which the potential crosses the horizontal axis. Above such critical separation, when the curve $V_{q\bar{q}}$ still exists, the preferred string configuration corresponds to non-interacting quarks whose strings fall on straight lines from infinity to the horizon. Such critical length is called the screening length. The configuration of interacting quarks in this case is a local minimum of the NG action, but does not dominate and corresponds to a metastable configuration. The upper branches never dominate, and they actually becomes a unstable---see e.g. \cite{Arias:2009me}. For separations above the one leading to the non-analyticity in the $V_{q\bar{q}}$ curve, there is no interaction at all between the quark-antiquark pair which have been completely screened by the thermal nature of the background. All of these potentials are non-confining, as expected in terms of general entropic arguments \cite{Maldacena:1998im,Witten:1998zw}.  

In the precise cases analyzed in this section, namely the Hayward, Dymnikova and Bardeen-like, planar, regular black holes, we observe that the value of the higher-curvature coupling $\alpha$ has an impact on the screening length. Remarkably, for Hayward and Dymnikova regular planar black holes, the screening length decreases as the coupling increases, while for the Bardeen case, we observe the opposite behavior. It is interesting to realize that for the latter case, the coupling $\alpha_2$ vanishes, while for the former two, the higher-curvature corrections to GR are closer to generic, since the series contains all the powers in the curvature, namely $\alpha_i\neq 0$ for all $i\geq 2$. In the holographic interpretation of the $\alpha'$ corrections in the bulk, the presence of those terms corresponds to finite t'Hooft coupling ($\lambda$) corrections. The behavior of the screening length for the first two cases, Hayward and Dymnikova black holes, are consistent with the fact that at finite t'Hooft coupling the interaction between the quark-antiquark pair decreases with respect to that in the strict $\lambda\rightarrow\infty$ limit, and the pair $q\bar{q}$ dissociates at shorter distances. A similar phenomenology occurs for the holographic thermalization of IR modes when $\alpha'^3$ corrections are included in Type IIB SUGRA, whose thermalization is delayed due to the presence of finite $\lambda$ corrections  \cite{Baron:2013cya}. It might be interesting to establish whether the behavior of the screening length as a function of the higher-curvature couplings could be used as a restriction criterion in a bottom-up approach to the higher-curvature corrections. We have consistently found that the Bardeen-like black holes backgrounds behave differently in what regards to the quark-antiquark potential, as it is the case for the thermodynamics of small black holes.

\section{Mapping Type-IIB effective action at $\mathcal O({\alpha'}^3)$ to QT gravity}\label{sec:6}
In a strict sense, the validity of the holographic interpretations provided above for the phase transitions, QN modes and string probes, would require the infinite series of QT terms to have a UV-complete, string theoretical origin. A complete top-down answer to this question is beyond the reach of current techniques of theoretical physics. Nevertheless, below we show that for Type-IIB SUGRA on $S^5$, the $\mathcal{\alpha'}^3$ correction which is quartic in the curvature, can be re-casted via field redefinition as a series of quadratic, cubic and quartic QT combinations. Actually, in \cite{Bueno:2019ltp}, it was shown that the gravitational sector of the ten-dimensional Type-IIB String Theory effective action on AdS$_5\times S^5$ truncated at (sub)leading order in $\alpha'$ can be mapped to a certain proper quartic GQTG---by proper, we mean that it has second order equations of motion after a trivial first integration, falling into the GQT family subclass but not in the QT gravity one. Parallelly, it was argued that the map could be done to a QT gravity but no explicit example was provided. Here we fill this gap, indicating the explicit field redefinition that maps Type-IIB effective action $\mathcal O({\alpha'}^3)$ to a QT gravity.

In this effective theory, the first stringy correction appears at ${\alpha'}^3$ order \cite{Gross:1986iv,Grisaru:1986px}, $I_{\text{IIB}}=I_{\text{IIB}}^{(0)}+{\alpha'}^3 I_{\text{IIB}}^{(1)}+\ldots$, where $I_{\text{IIB}}^{(0)}$ is the usual two-derivative SUGRA action  \cite{Howe:1983sra}, and the dots stand for subleading corrections in  $\alpha'$. When the theory is defined on $\mathcal{A}_5\times S^5$ with $\mathcal{A}_5$ a negatively curved Einstein manifold, it is consistent to truncate all fields except the metric, allowing for the construction of an effective action for the five-dimensional metric \cite{Myers:2008yi,Buchel:2008ae,Galante:2013wta}, reading \cite{Gubser:1998nz,Buchel:2004di}
\begin{equation}\label{acads5}
I_{\text{IIB}_{{\mathcal{A}}_5 \times S^5}}=\frac{1}{16\pi G}\int\diff^5 x\sqrt{|g|}\left[ R+\frac{12}{L^2}+\frac{\zeta(3)}{8}{\alpha'}^3 W^4 \right] \, ,\quad 
\end{equation}
with $W^4=\left(W_{abcd}W^{ebcf}+\frac{1}{2}W_{adbc}W^{efbc}\right)W\indices{^a^g_h_e}W\indices{_f_g^h^d}$.

The field redefinition introduced in \cite{Bueno:2019ltp}, which completed the quartic-order term $W^4$ appearing in the action \eqref{acads5} into a GQTG, involved a rank-two symmetric tensor cubic in curvature, $\hat{C}_{ab}^{(3)}$, constructed with a specific choice of internal coefficients and contracted with a Ricci tensor. When a cosmological constant is included, the redefinition produces an additional cubic term in the action, proportional to the trace $\hat{C} = g^{ab} \hat{C}_{ab}^{(3)}$, where $g^{ab}$ is the inverse of the redefined metric. Demanding that this trace term $\hat{C}$ also belongs to the GQT class imposes further constraints on the remaining coefficients, preventing the construction of a QT theory via this approach alone.

As anticipated in \cite{Bueno:2019ltp}, completing an action to a QT theory in the presence of a cosmological constant requires a field redefinition that includes contributions from multiple orders in curvature simultaneously. This is precisely what we illustrate here. To that end, consider the redefinition 
\begin{equation}
g_{ab} \rightarrow g_{ab} - \frac{\zeta(3)}{8}\alpha'^3 \left[\frac{1}{L^4}\left(\hat{C}^{(1)}_{ab} - \frac{g_{ab}}{3} \hat{C}^{(1)} \right)+\frac{1}{L^2}\left(\hat{C}^{(2)}_{ab} - \frac{g_{ab}}{3} \hat{C}^{(2)} \right)+\hat{C}^{(3)}_{ab} - \frac{g_{ab}}{3} \hat{C}^{(3)}\right]\,,
\label{metredefi}
\end{equation}
where the rank-two tensors of order $i$ in curvature, $\hat C^{(i)}_{ab}$, read explicitly
\begin{align}
\hat C_{ab}^{(3)}=&-\frac{6935}{1584}\tensor{R}{^{cdef}}\tensor{R}{_c^g_{ea}}\tensor{R}{_{dgfb}}-\frac{18625}{19008}\tensor{R}{^{cdef}}\tensor{R}{_{cd}^g_a}\tensor{R}{_{efgb}}+\frac{7}{44}\tensor{R}{_a^c_b^d} \tensor{R}{_{efgc}} \tensor{R}{^{efg}_{d}} \\\notag &+\frac{2053}{1584} g_{ab} \tensor{R}{_g^c_h^d} \tensor{R}{_c^e_d^f}\tensor{R}{_e^g_f^h}+\frac{1241}{19008}g_{ab}\tensor{R}{_{gh}^{cd}}\tensor{R}{_{cd}^{ef}}\tensor{R}{_{ef}^{gh}}+\frac{6035}{3168}R^{cd}\tensor{R}{^{e}_a^f_c}\tensor{R}{_{ebfd}}\\ \notag &  
+  \frac{24215}{6336} R^{cd}\tensor{R}{^{e}_a^f_b}\tensor{R}{_{ecfd}}
-\frac{1291}{12672}R^{cd}\tensor{R}{^{ef}_{ac}}\tensor{R}{_{efbd}} -\frac{767}{1584}R_{ab}R_{cdef}R^{cdef}\\ \notag &  +\frac{2155}{12672}g_{ab} R_{ghcd} \tensor{R}{^{ghc}_{e}}R^{de}-\frac{521}{1408}R R_{ghca} \tensor{R}{^{ghc}_{b}}+\frac{1679}{12672}g_{ab}R R_{ghcd}R^{ghcd}\\ \notag & 
-\frac{25}{44}R^{gh}R_{gahd} R_{b}^d-\frac{3569}{3168} R R_{gacb}R^{gc}+ \frac{7}{704} g_{ab}R^3\, .\\
\hat C_{ab}^{(2)}=&\frac{15}{44}\tensor{R}{_{acde}}\tensor{R}{_b^{cde}}-\frac{285 }{176}\tensor{R}{_{cd}}\tensor{R}{_a^c_b^d}+\frac{15}{44}R\tensor{R}{_{ab}}\, ,\quad \hat C_{ab}^{(1)}=\frac{45}{44}R_{ab}\,.
\end{align}
After applying this redefinition in action \eqref{acads5}, we complete the $W^4$ term to a QT, and QT terms of a lower order appear

\begin{align}\label{tilde2I}\nonumber
\tilde I=&\frac{1}{16\pi G}\int\diff^5 x\sqrt{|g|}\left(\frac{12}{L^2}+R+\frac{15\zeta(3)}{88}{\alpha'}^3\bigg[\frac{3}{L^6}R+\frac{1}{L^4}\left(R_{abcd}R^{abcd}-4R_{ab}R^{ab}+R^2\right)\right.\\
&\left.+\frac{1}{L^2}Q^{(3)}\left(\mathcal{R}^3\right)+Q^{(4)}\left(\mathcal{R}^4\right)\bigg]\right) \, ,
\end{align}
with the QT combinations $Q^{(3)}$ and $Q^{(4)}$ given by
\begin{align}
Q^{(3)}\left(\mathcal{R}^3\right)&=\frac{11}{15}\left(4\hat C^{(3)}+R^{ab}\hat C^{(2)}_{ab}\right)\ ,\\
Q^{(4)}\left(\mathcal{R}^4\right)&=\frac{11}{15}\left(W^4+R^{ab}\hat{C}_{ab}^{(3)}\right)\ .
\end{align}
With this result at hand, it can be easily checked that the action \eqref{tilde2I} is indeed of a QT family since the integrated equations of motion for $f(r)$ lead to a polynomial equation on such function, namely
% \begin{equation}
% \frac{1}{L^2} + \psi +\frac{15 \zeta(3)}{88} {\alpha’}^3 \left( 3\frac{\psi}{L^6} + 2\frac{\psi^2}{L^4} -\frac{4}{3}\frac{\psi^3}{L^2} - \psi^4 \right) = \frac{m}{r^4}\,,
% \end{equation}
\begin{equation}
\frac{r^2}{L^2} +\left(k-f\right)+\frac{15 \zeta(3)\alpha’^3}{88}  \left(\frac{3(k-f)}{L^6}+\frac{2(k-f)^2}{r^2L^4} -\frac{4(k-f)^3}{3r^4L^2} - \frac{(k-f)^4}{r^6} \right) = \frac{m}{r^2}\,,
\end{equation}
for the metric function $f=f(r)$, where $m$ is an integration constant. The result displays, as expected, the polynomial behavior characteristic of the QT subclass. Notice that the field redefinition has led to a perturbative contribution to Newton's constant coming from the term proportional to $\alpha'^3L^{-6}$. Notice that the sign of the Gauss-Bonnet term that emerged from the field redefinition is positive, as it is the case in other string theory frameworks that lead to a perturbative Gauss-Bonnet term. 

\section{Generalized First Law and Smarr formula}\label{sec:7}
Smarr Formula for asymptotically flat black holes in GR, provides a relation between the mass, temperature and entropy, namely $(D-3)M=(D-2)TS$, and emerges from a natural scaling argument from the first law of black hole thermodynamics. The dependence on the dimension appears due to the fact that in such setup the entropy is given by the horizon area, and not in terms of the volume, as is the case for a standard thermal system. This formula stands for the gravitational version of the Euler relation in Statistical Physics, which is valid for homogeneous systems and can be deduced in any ensemble. It is well-known that in the presence of a negative cosmological constant such relation ceases to be valid. The inclusion of a new, dimensionful quantity precludes the use of the standard scaling argument from the first law of black hole thermodynamics. In \cite{Kastor:2009wy,Dolan:2010ha}, it was realized, and further developed, that extending the first law by the inclusion of a work term of the form $V\diff P$ allows recovering a finite relation between thermodynamic quantities, provided $P$ is identified with the cosmological constant as $P\sim-\Lambda$ and the conjugate quantity $V$ represents a thermodynamic volume, opening the rich field of Black Hole Chemistry (see \cite{Kubiznak:2016qmn} and references therein\footnote{For a holographic interpretation of Black Hole Chemistry see \cite{Ahmed:2023snm,Ahmed:2023dnh} and for a recent covariant method that upgrades coupling constants to fields, permitting to interpret Black Hole Chemistry as standard black hole thermodynamics see \cite{Hajian:2023bhq}.}). In this framework, the mass of the black holes turns out to be a function of $M=M(S,Q,P)$---$Q$ standing for other global charges involved, and in consequence it must be interpreted as the enthalpy instead of the internal energy of the black hole, since the control variable is the pressure and not the volume. The inclusion of higher-curvature terms with dimensionful couplings also requires the introduction of new chemical potentials at the level of the first law, if one is willing to keep the validity of an Euler relation. In Lovelock gravity this was originally formalized in \cite{Kastor:2010gq}---see also \cite{Cai:2013qga} and for black strings \cite{Henriquez-Baez:2022bfi}---and further applied in QT gravity \cite{Hennigar:2015esa}, where the extension of the first law and Smarr formula works provided one includes variations of the coupling $\alpha$. Here we show that these ideas naturally extend for the gravitational theories that admit regular black holes. The precise form of the extended first law would depend on the number of couplings. 

The structure of the regular black holes studied in this work, depending on the choice of the couplings $\alpha_i$, is captured by the function $h(\psi)$. Let us first prove that imposing the standard Smarr relation $(D-3)M=(D-2)TS$ implies that $h(\psi)\propto \psi$, leading to GR without a cosmological constant. The proof is general and proceeds as follows. We first perform a partial integration in the expression for the entropy given in Section~\ref{sec:2}, obtaining
%
%\[M=\frac{(D-2)\Sigma_{D-2,k}}{16\pi G}r_+^{D-1}h(\psi_+)\]
%\[T=\frac 1{4\pi}\,\frac1{r_+}\left((D-1)\frac{r_+^{2}h(\psi_+)}{h'(\psi_+)}-2\right)\]
%\[S=-\frac{(D-2)\Sigma_{D-2,k}}{8 G}\int h'(\psi_+)\frac{d\psi_+}{\psi_+^{D/2}}\]
%A partial integration of the last equation gives
%
\begin{equation}
S=-\frac{(D-2)\Sigma_{D-2,k}}{8 G}\left(\frac{h(\psi_+)}{\psi_+^{D/2}}-\int h(\psi_+)\,\diff\left(\frac{1}{\psi_+^{D/2}}\right)\right)\,.
\label{entropyx}
\end{equation}
The form of the integral inspires the definition of a new variable $t=1/\psi^{D/2}=r^D_+$,
which can then be inserted into the expression \eqref{entropyx} as well as in those for energy and temperature. Using the transformation $h'=-\frac D2\frac {\dot h}{\psi^{D/2+1}}=-\frac D2\dot h\,t^{1+\frac2D}$ we obtained
\begin{eqnarray}
M&=&\frac{(D-2)\Sigma_{D-2,k}}{16\pi G}\,t^{1-\frac 1D}\,h(t)\,,\\
T&=&\frac 1{4\pi}\,\frac1{t^{\frac1D}}\left(-\frac{2(D-1)}D\frac{h(t)}{\dot h(t)\,t}-2\right)\,,\\
S&=&-\frac{(D-2)\Sigma_{D-2,k}}{8 G}\left({h(t)}\,t-\int h(t)\,\diff t\right)\,.
\end{eqnarray}
These expressions get simpler by writing them in terms of the primitive function $H(t)$ such that $\dot H(t)=h(t)$, as
\begin{eqnarray}
M&=&\frac{(D-2)\Sigma_{D-2,k}}{16\pi G}\,t^{1-\frac 1D}\,\dot H(t)\,,\\
T&=&-\frac 1{2\pi}\,\frac 1{t^{\frac1D}}\left(\frac{(D-1)}D\frac{\dot H(t)}{t\ddot H(t)}+1\right)\,,\\
S&=&-\frac{(D-2)\Sigma_{D-2,k}}{8 G}\left({\dot H(t)}\,t-H(t)\right)\,.
\end{eqnarray}
Now we can impose the standard Smarr relation $(D-3)M=(D-2)TS$ as a constraint in the function $H$,
resulting in the ordinary differential equation 

\begin{equation}
D\,t^2\ddot H(t)\dot H(t)+(D-2)\left({(D-1)}{\dot H(t)}\left({\dot H(t)}\,t-H(t)\right)-D\,t\,\ddot H(t)\,H(t)\right)=0\,.
\end{equation}
This equation can be solved with a power law ansatz $H \propto t^\chi$ where $\chi$ is a parameter to be determined. We obtain the trivial solutions $\chi=0$ and $\chi=1$, and a non-trivial one $\chi=(D-2)/D$. This last form implies $h=\dot H\propto  t^{-2/D}$ or in other words $h\propto\psi$, proving that we only get the standard Smarr relation in the GR case. In consequence, the validity of a relation between finite thermodynamical variables, including a cosmological term and the higher-curvature terms, requires promoting the analysis to the one of generalized black hole thermodynamics. For concreteness, having in mind the Hayward, Dymnikova and Bardeen-like black holes, let us focus on higher-curvature terms controlled by a single new coupling $\alpha$. Considering the thermodynamic quantities $M,T$ and $S$ given in \eqref{eq:thermoM}, \eqref{eq:thermoT} and \eqref{eq:thermoS}, respectively, as functions of $\left(r_+,\Lambda,\alpha\right)$, one can prove that
\begin{equation}
\diff M=T\diff S+V\diff P+\mu_\alpha \diff \alpha\,,
\end{equation}
with
\begin{eqnarray}
P&=&-\frac{\Lambda}{8\pi G}=\frac{(D-1)(D-2)}{16\pi G L^2}\ ,\\
V&=&\frac{\Sigma_{D-2,k}}{(D-1)}r_+^{D-1}\ ,\\
\mu_\alpha&=& \frac{(D-2)\Sigma_{D-2,k}}{16\pi G} \left[ r_+^{D-1} \left. \frac{\partial h(\psi_+)}{\partial \alpha}\right|_{L,r_+} + \right. \nonumber \\
& & \ \ \ \ \ \ \ \ \ \ \ \ \ \ \ \ \ \ \left. \frac{k^{D/2-1}}{2\pi r_+}\left( \frac{(D-1)r_+^2h(\psi_+)}{h'(\psi_+)}-2k \right)\int \frac{\diff \psi_+}{\psi_+^{D/2}} \left. \frac{\partial h'(\psi_+)}{\partial \alpha} \right|_{L,r_+}
 \right]\, .
\end{eqnarray}
Notice that the expression for the volume and pressure take the same form as in GR, in terms of the horizon radius, while the variable conjugated to the coupling $\alpha$, depends on the theory via the function $h(\psi)$. With these equations one can prove that the following generalized Smarr formula is fulfilled
\begin{equation}\label{smarrfull}
(D-3)M=(D-2)TS-2VP+2\mu_\alpha\alpha\,.
\end{equation}
The analysis of phase transitions in the context of generalized thermodynamics is beyond the scope of the current work, but let us close this section mentioning that in GR, this framework has been useful to discover new phase transitions occurring at fixed pressure and therefore for fixed $\Lambda$---see \cite{Kubiznak:2016qmn}.

\section{Conclusions}\label{sec:8}
In this paper we have explored holographically inspired properties of regular black holes in pure gravity theories. These black holes extend to the topological and asymptotically AdS case, the asymptotically flat black holes recently constructed in \cite{Bueno:2024dgm}. Even though these black holes exist in any dimension, we have focus on the five-dimensional case\footnote{Following an agreement with the authors of \cite{Hennigar:2025ftm}, our manuscripts appeared the same day in the \texttt{ArXiv}. It is very interesting to notice that such authors explored the thermal properties of regular black holes in AdS in arbitrary dimension, and found a new type of spherical black hole-black hole phase transition in dimension $9$ for the Hayward-like family.}. The black holes are supported by an infinite series of quasitopological terms, containing higher powers in the curvature of arbitrary order. These terms exist in any dimension and possess some remarkable properties: they lead to higher-curvature gravity theories fulfilling a Birkhoff's theorem, they allow for the construction of holographic c-functions and around the maximally AdS solution they lead to stable perturbations since the graviton propagator coincides with that of GR, with an effective Newton's constant. In Section 6, we presented the explicit field redefinition that allows recasting the $\mathcal{O}(\alpha'^3)$, pure gravity correction of Type IIB SUGRA on $\mathcal{A}_5\times S^5$, as a series of quadratic, cubic and quartic quasitopological terms. The properties of the latter allow to stablish at a perturbative level in $\alpha'$, the perturbative stability of the AdS background. It was proved in \cite{Bueno:2019ltp} that any higher-curvature gravity theory, algebraic in the Riemann tensor, can be recasted as a GQTG gravity, and it was also argued in such reference that a field redefinition to strict QT gravities does exist in such frameworks. It would be interesting to explore whether still including the $\alpha'$ corrections containing the Ramond-Ramond five-form of Type IIB SUGRA one can obtain a sensible theory order by order in perturbation theory (see e.g. \cite{Paulos:2008tn} and \cite{Pawelczyk:1998pb}). The inclusion of these terms in the realm of AdS/CFT stands for considering the physics at finite 't Hooft coupling.

Allowing for the presence of an infinite series of quasitopological terms, in a bottom-up approach, we studied the thermodynamics of the regular asymptotically AdS black holes, at fixed temperature. The inclusion of the dimensionful coupling $\alpha$ may lead to three families of black holes for a given temperature. The small black holes are always continuously connected to an extremal one, which has a minimum radius for finite $\alpha$, and therefore they cannot be arbitrarily small. These black holes have positive heat capacity. The intermediate black holes on the contrary, have negative heat capacity. Finally, there is a branch of large regular black holes that have positive heat capacity, and as in GR above a given temperature, they dominate the canonical ensemble since there is a first-order phase transition with thermal AdS. We studied the effect of the higher-curvature couplings on the phase transition temperature, and we showed that the onset of the transition moves towards lower temperature for the Hayward and Dymnikova regular black holes, while it moves towards higher temperatures for the Bardeen-like black hole. In the same cases, we computed the behavior of the fundamental mode of a massless scalar probe as a function of the temperature. For large black holes, as the temperature increases, the lifetime of the longest-lived mode is reduced via a linear relation. This has a clear parallel with the behavior of conductivity in a metal. The behavior of the real part of the frequencies, for large black holes also has a natural interpretation. We proved that for the three mentioned families of large black holes the temperature increases linearly with the black hole radius, therefore, for such branch of black holes as the temperature increases, a sensible notion of renormalized volume of the black hole exterior must decrease. We observe that the real part of the frequency increases with the temperature of large black holes, and therefore the ``pitch" increases as the exterior volume decreases, which has a clear parallel with the sound of the drum. In the ingoing-null time $v$, the quasinormal frequencies we obtained fulfill analyticity at the horizon and vanish at infinite, namely they vanish at the boundary of the drum. It would be interesting to explore the behavior of quasinormal frequencies for massive scalar probes with more general boundary conditions in the range of negative squared mass, above the Breitenlohner-Freedman bound \cite{Breitenlohner:1982bm,Breitenlohner:1982jf}. The last holographic probe we studied on the asymptotically AdS regular black holes was a U-shaped string attached to the boundary, whose on-shell Nambu-Goto action allows to read the quark-antiquark potential in the dual field theory. As for black hole thermodynamics, the behavior of the screening length for the planar Hayward and Dymnikova black holes is different than the planar Bardeen case. In the latter, the screening length increases with the higher-curvature coupling (finite, large 't Hooft coupling corrections), while in the former two cases the quarks dissociate earlier as $\alpha$ increases\footnote{For special values of the couplings the inclusion of higher-curvature terms may allow for black holes in vacuum, and for the behavior of strings attached to the boundaries of such spacetimes with non-trivial topology, see \cite{Chernicoff:2020tvr,Chernicoff:2022slp,Arias:2010xg,Lu:2013awa}.}. It has been recently argued that U-shaped string probes can capture the presence of phase transitions between different plasma phases in the dual theory \cite{Anabalon:2025uzl,Anabalon:2024lgp}. Whether such property extends to the presence of higher-curvature terms on regular black holes is a project left for the future, nevertheless it is important mentioning that the realm of quasitopological gravity defines a perfect scenario to explore these ideas due to the computational control over the thermodynamics and the construction of exact solutions. It is interesting mentioning that the inclusion of a scalar field coupled to higher-curvature terms still leading to second order field equations and a Birkhoff's-like theorem on spherical symmetry, allows constructing regular black holes in four and three dimensions \cite{Fernandes:2025fnz,Fernandes:2025eoc,Bueno:2025jgc,Cisterna:2025vxk}, and given the relevance of CFTs in dimensions three and two, respectively, it would be interesting to explore the holographic probes considered in this work, in such scenarios.

Let us finish the conclusions with some comments about the covariant structure of the quasitopological theories admitting regular black holes. For the sake of the argument that follows, let us denote by $\mathcal{Z}_0$ and $\mathcal{Z}_1$ the cosmological term and the Einstein-Hilbert Lagrangian, denoted as the zeroth-order and first-order quasitopological Lagrangians. Consequently $\mathcal{Z}_2$ is the Gauss-Bonnet combination, quadratic in the Riemann tensor and $\mathcal{Z}_3$, $\mathcal{Z}_4$ and $\mathcal{Z}_5$ denote the cubic, quartic and quintic in the Riemann, quasitopological combinations. In \cite{Bueno:2019ycr} it was shown that there is a precise, non-linear recurrence relation that allows, from the first five quasitopological Lagrangians, constructing a whole series of them, at arbitrary order, namely for $n\geq 6$, one has that $\mathcal{Z}_n=\mathcal{Z}_n(\mathcal{Z}_0,...,\mathcal{Z}_5)$. When this recurrence relation is implemented at the level of the Lagrangian, and the couplings are fixed in the different manners that lead to the regular black holes controlled by the analytic function $h(\psi)$, the following resumation at the level of the action must occur
\begin{align}
I&=\int\sqrt{-g}d^Dx\left(R-2\Lambda+\sum_{n=2}^\infty\alpha_n\mathcal{Z}_n\right)\\
&=\int\sqrt{-g}d^Dx\left(R-2\Lambda+F\left(R,\mathcal{Z}_2,...,\mathcal{Z}_5\right)\right)\,,
\end{align}
for some function $F$ that depends on the choice of the couplings $\alpha_n$, namely on the function $h(\psi)$. In order to avoid the resummation of the non-linear recurrence relation one could actually proceed in the opossite direction, namely to find the function $F$ of the first five QT combinations that leads to the Hayward, Dymnikova or Bardeen-like regular black holes. Once the function $F$ is fixed, leading to a non-perturbative QT gravity fulfilling all the structural properties that these theories posses, one can start exploring the theory beyond spherical symmetry, for finite $\alpha$. We expect to report on this approach in the future, but notice that this approach will produce a particular family of QT lagrangians.  While QT combinations of a given order coincide on-shell on spherical symmetry, already at the slowly rotating level their predictions may significantly differ (see e.g. equation 29 of \cite{Fierro:2020wps} for four types of contributions emerging on quartic QT gravities, on slowly rotating black holes). 

\section*{Acknowledgments}

We thank Felipe Agurto, Andrés Anabalón, Pablo Bueno, Pablo A. Cano, Mariano Chernicoff, Adolfo Cisterna, Alberto Faraggi, Gaston Giribet, Robie A. Hennigar and Ángel J. Murcia, Marcelo Oyarzo and Guillermo Silva for useful discussions. M.A. is supported by ANID Doctoral Fellowship 21252777. L. G. is supported by ANID Master Fellowship 1047241. The work of JM is supported by ANID FONDECYT Postdoctorado Grant No. 3230626. JO is supported by FONDECYT grant 1221504.

\appendix

\bibliographystyle{JHEP-2}
\bibliography{Gravities.bib}

\providecommand{\href}[2]{#2}\begingroup\raggedright\begin{thebibliography}{100}

\bibitem{Bueno:2024dgm}
P.~Bueno, P.~A. Cano and R.~A. Hennigar, {\it {Regular black holes from pure
  gravity}},  {\em Phys. Lett. B} {\bf 861} (2025) 139260
  [\href{http://arXiv.org/abs/2403.04827}{{\tt 2403.04827}}].

\bibitem{Ayon-Beato:1998hmi}
E.~Ayon-Beato and A.~Garcia, {\it {Regular black hole in general relativity
  coupled to nonlinear electrodynamics}},  {\em Phys. Rev. Lett.} {\bf 80}
  (1998) 5056--5059 [\href{http://arXiv.org/abs/gr-qc/9911046}{{\tt
  gr-qc/9911046}}].

\bibitem{Cisterna:2020rkc}
A.~Cisterna, G.~Giribet, J.~Oliva and K.~Pallikaris, {\it {Quasitopological
  electromagnetism and black holes}},  {\em Phys. Rev. D} {\bf 101} (2020),
  no.~12 124041 [\href{http://arXiv.org/abs/2004.05474}{{\tt 2004.05474}}].

\bibitem{Witten:1998zw}
E.~Witten, {\it {Anti-de Sitter space, thermal phase transition, and
  confinement in gauge theories}},  {\em Adv. Theor. Math. Phys.} {\bf 2}
  (1998) 505--532 [\href{http://arXiv.org/abs/hep-th/9803131}{{\tt
  hep-th/9803131}}].
%%CITATION = HEP-TH/9803131;%%

\bibitem{Horowitz:1999jd}
G.~T. Horowitz and V.~E. Hubeny, {\it {Quasinormal modes of AdS black holes and
  the approach to thermal equilibrium}},  {\em Phys. Rev. D} {\bf 62} (2000)
  024027 [\href{http://arXiv.org/abs/hep-th/9909056}{{\tt hep-th/9909056}}].

\bibitem{Maldacena:1998im}
J.~M. Maldacena, {\it {Wilson loops in large N field theories}},  {\em Phys.
  Rev. Lett.} {\bf 80} (1998) 4859--4862
  [\href{http://arXiv.org/abs/hep-th/9803002}{{\tt hep-th/9803002}}].

\bibitem{Oliva:2010eb}
J.~Oliva and S.~Ray, {\it {A new cubic theory of gravity in five dimensions:
  Black hole, Birkhoff's theorem and C-function}},  {\em Class. Quant. Grav.}
  {\bf 27} (2010) 225002 [\href{http://arXiv.org/abs/1003.4773}{{\tt
  1003.4773}}].
%%CITATION = ARXIV:1003.4773;%%

\bibitem{Myers:2010ru}
R.~C. Myers and B.~Robinson, {\it {Black Holes in Quasi-topological Gravity}},
  {\em JHEP} {\bf 08} (2010) 067 [\href{http://arXiv.org/abs/1003.5357}{{\tt
  1003.5357}}].
%%CITATION = ARXIV:1003.5357;%%

\bibitem{Dehghani:2011vu}
M.~H. Dehghani, A.~Bazrafshan, R.~B. Mann, M.~R. Mehdizadeh, M.~Ghanaatian and
  M.~H. Vahidinia, {\it {Black Holes in Quartic Quasitopological Gravity}},
  {\em Phys. Rev.} {\bf D85} (2012) 104009
  [\href{http://arXiv.org/abs/1109.4708}{{\tt 1109.4708}}].
%%CITATION = ARXIV:1109.4708;%%

\bibitem{Cisterna:2017umf}
A.~Cisterna, L.~Guajardo, M.~Hassaine and J.~Oliva, {\it {Quintic
  quasi-topological gravity}},  {\em JHEP} {\bf 04} (2017) 066
  [\href{http://arXiv.org/abs/1702.04676}{{\tt 1702.04676}}].
%%CITATION = ARXIV:1702.04676;%%

\bibitem{Bueno:2019ycr}
P.~Bueno, P.~A. Cano and R.~A. Hennigar, {\it {(Generalized) quasi-topological
  gravities at all orders}},  {\em Class. Quant. Grav.} {\bf 37} (2020), no.~1
  015002 [\href{http://arXiv.org/abs/1909.07983}{{\tt 1909.07983}}].

\bibitem{Moreno:2023rfl}
J.~Moreno and A.~J. Murcia, {\it {Classification of generalized
  quasitopological gravities}},  {\em Phys. Rev. D} {\bf 108} (2023), no.~4
  044016 [\href{http://arXiv.org/abs/2304.08510}{{\tt 2304.08510}}].

\bibitem{Ahmed:2017jod}
J.~Ahmed, R.~A. Hennigar, R.~B. Mann and M.~Mir, {\it {Quintessential Quartic
  Quasi-topological Quartet}},  {\em JHEP} {\bf 05} (2017) 134
  [\href{http://arXiv.org/abs/1703.11007}{{\tt 1703.11007}}].
%%CITATION = ARXIV:1703.11007;%%

\bibitem{Bueno:2022res}
P.~Bueno, P.~A. Cano, R.~A. Hennigar, M.~Lu and J.~Moreno, {\it {Generalized
  quasi-topological gravities: the whole shebang}},  {\em Class. Quant. Grav.}
  {\bf 40} (2023), no.~1 015004 [\href{http://arXiv.org/abs/2203.05589}{{\tt
  2203.05589}}].

\bibitem{Moreno:2023arp}
J.~Moreno and A.~J. Murcia, {\it {Cosmological higher-curvature gravities}},
  {\em Class. Quant. Grav.} {\bf 41} (2024), no.~13 135017
  [\href{http://arXiv.org/abs/2311.12104}{{\tt 2311.12104}}].

\bibitem{Bueno:2016xff}
P.~Bueno and P.~A. Cano, {\it {Einsteinian cubic gravity}},  {\em Phys. Rev.}
  {\bf D94} (2016), no.~10 104005 [\href{http://arXiv.org/abs/1607.06463}{{\tt
  1607.06463}}].
%%CITATION = ARXIV:1607.06463;%%

\bibitem{Hennigar:2016gkm}
R.~A. Hennigar and R.~B. Mann, {\it {Black holes in Einsteinian cubic
  gravity}},  {\em Phys. Rev.} {\bf D95} (2017), no.~6 064055
  [\href{http://arXiv.org/abs/1610.06675}{{\tt 1610.06675}}].
%%CITATION = ARXIV:1610.06675;%%

\bibitem{Bueno:2016lrh}
P.~Bueno and P.~A. Cano, {\it {Four-dimensional black holes in Einsteinian
  cubic gravity}},  {\em Phys. Rev.} {\bf D94} (2016), no.~12 124051
  [\href{http://arXiv.org/abs/1610.08019}{{\tt 1610.08019}}].
%%CITATION = ARXIV:1610.08019;%%

\bibitem{lovelock1970divergence}
D.~Lovelock, {\it Divergence-free tensorial concomitants},  {\em aequationes
  mathematicae} {\bf 4} (1970), no.~1 127--138.

\bibitem{Lovelock:1971yv}
D.~Lovelock, {\it {The Einstein tensor and its generalizations}},  {\em J.
  Math. Phys.} {\bf 12} (1971) 498--501.
%%CITATION = JMAPA,12,498;%%

\bibitem{Wheeler:1985nh}
J.~T. Wheeler, {\it {Symmetric Solutions to the Gauss-Bonnet Extended Einstein
  Equations}},  {\em Nucl. Phys.} {\bf B268} (1986) 737--746.
%%CITATION = NUPHA,B268,737;%%

\bibitem{Boulware:1985wk}
D.~G. Boulware and S.~Deser, {\it {String Generated Gravity Models}},  {\em
  Phys. Rev. Lett.} {\bf 55} (1985) 2656.
%%CITATION = PRLTA,55,2656;%%

\bibitem{Cai:2001dz}
R.-G. Cai, {\it {Gauss-Bonnet black holes in AdS spaces}},  {\em Phys. Rev.}
  {\bf D65} (2002) 084014 [\href{http://arXiv.org/abs/hep-th/0109133}{{\tt
  hep-th/0109133}}].
%%CITATION = HEP-TH/0109133;%%

\bibitem{Padmanabhan:2013xyr}
T.~Padmanabhan and D.~Kothawala, {\it {Lanczos-Lovelock models of gravity}},
  {\em Phys. Rept.} {\bf 531} (2013) 115--171
  [\href{http://arXiv.org/abs/1302.2151}{{\tt 1302.2151}}].
%%CITATION = ARXIV:1302.2151;%%

\bibitem{Corral:2019leh}
C.~Corral, D.~Flores-Alfonso and H.~Quevedo, {\it {Charged Taub-NUT solution in
  Lovelock gravity with generalized Wheeler polynomials}},  {\em Phys. Rev. D}
  {\bf 100} (2019), no.~6 064051 [\href{http://arXiv.org/abs/1908.06908}{{\tt
  1908.06908}}].

\bibitem{Bueno:2018uoy}
P.~Bueno, P.~A. Cano, R.~A. Hennigar and R.~B. Mann, {\it {NUTs and bolts
  beyond Lovelock}},  {\em JHEP} {\bf 10} (2018) 095
  [\href{http://arXiv.org/abs/1808.01671}{{\tt 1808.01671}}].
%%CITATION = ARXIV:1808.01671;%%

\bibitem{Corral:2022udb}
C.~Corral, D.~Flores-Alfonso, G.~Giribet and J.~Oliva, {\it {Higher-curvature
  generalization of Eguchi-Hanson spaces}},  {\em Phys. Rev. D} {\bf 106}
  (2022), no.~8 084055 [\href{http://arXiv.org/abs/2207.04014}{{\tt
  2207.04014}}].

\bibitem{Corral:2025yvr}
C.~Corral, B.~Diez, D.~Flores-Alfonso, N.~Merino and L.~Sanhueza, {\it
  {Inhomogeneous metrics on complex bundles in Lovelock gravity}},
  \href{http://arXiv.org/abs/2504.11562}{{\tt 2504.11562}}.

\bibitem{Myers:2010jv}
R.~C. Myers, M.~F. Paulos and A.~Sinha, {\it {Holographic studies of
  quasi-topological gravity}},  {\em JHEP} {\bf 08} (2010) 035
  [\href{http://arXiv.org/abs/1004.2055}{{\tt 1004.2055}}].
%%CITATION = ARXIV:1004.2055;%%

\bibitem{Bueno:2018xqc}
P.~Bueno, P.~A. Cano and A.~Ruip\'erez, {\it {Holographic studies of
  Einsteinian cubic gravity}},  {\em JHEP} {\bf 03} (2018) 150
  [\href{http://arXiv.org/abs/1802.00018}{{\tt 1802.00018}}].
%%CITATION = ARXIV:1802.00018;%%

\bibitem{Li:2018drw}
Y.-Z. Li, H.~Lü and Z.-F. Mai, {\it {Universal Structure of Covariant
  Holographic Two-Point Functions In Massless Higher-Order Gravities}},  {\em
  JHEP} {\bf 10} (2018) 063 [\href{http://arXiv.org/abs/1808.00494}{{\tt
  1808.00494}}].
%%CITATION = ARXIV:1808.00494;%%

\bibitem{Cano:2022ord}
P.~A. Cano, A.~J. Murcia, A.~Rivadulla~S\'anchez and X.~Zhang, {\it
  {Higher-derivative holography with a chemical potential}},  {\em JHEP} {\bf
  07} (2022) 010 [\href{http://arXiv.org/abs/2202.10473}{{\tt 2202.10473}}].

\bibitem{Bueno:2018yzo}
P.~Bueno, P.~A. Cano, R.~A. Hennigar and R.~B. Mann, {\it {Universality of
  Squashed-Sphere Partition Functions}},  {\em Phys. Rev. Lett.} {\bf 122}
  (2019), no.~7 071602 [\href{http://arXiv.org/abs/1808.02052}{{\tt
  1808.02052}}].
%%CITATION = ARXIV:1808.02052;%%

\bibitem{Bueno:2020odt}
P.~Bueno, P.~A. Cano, R.~A. Hennigar, V.~A. Penas and A.~Ruip\'erez, {\it
  {Partition functions on slightly squashed spheres and flux parameters}},
  {\em JHEP} {\bf 04} (2020) 123 [\href{http://arXiv.org/abs/2001.10020}{{\tt
  2001.10020}}].

\bibitem{Bueno:2022jbl}
P.~Bueno, P.~A. Cano, A.~Murcia and A.~Rivadulla~S\'anchez, {\it {Universal
  Feature of Charged Entanglement Entropy}},  {\em Phys. Rev. Lett.} {\bf 129}
  (2022), no.~2 021601 [\href{http://arXiv.org/abs/2203.04325}{{\tt
  2203.04325}}].

\bibitem{Dey:2016pei}
A.~Dey, P.~Roy and T.~Sarkar, {\it {On holographic R\'enyi entropy in some
  modified theories of gravity}},  {\em JHEP} {\bf 04} (2018) 098
  [\href{http://arXiv.org/abs/1609.02290}{{\tt 1609.02290}}].

\bibitem{Bueno:2020uxs}
P.~Bueno, J.~Camps and A.~V. L\'opez, {\it {Holographic entanglement entropy
  for perturbative higher-curvature gravities}},  {\em JHEP} {\bf 04} (2021)
  145 [\href{http://arXiv.org/abs/2012.14033}{{\tt 2012.14033}}].

\bibitem{Caceres:2020jrf}
E.~C\'aceres, R.~C. V\'asquez and A.~Vilar~L\'opez, {\it {Entanglement entropy
  in cubic gravitational theories}},  {\em JHEP} {\bf 05} (2021) 186
  [\href{http://arXiv.org/abs/2009.11595}{{\tt 2009.11595}}].

\bibitem{Anastasiou:2022pzm}
G.~Anastasiou, I.~J. Araya, A.~Argando\~na and R.~Olea, {\it {CFT correlators
  from shape deformations in Cubic Curvature Gravity}},  {\em JHEP} {\bf 11}
  (2022) 031 [\href{http://arXiv.org/abs/2208.00093}{{\tt 2208.00093}}].

\bibitem{Mir:2019ecg}
M.~Mir, R.~A. Hennigar, J.~Ahmed and R.~B. Mann, {\it {Black hole chemistry and
  holography in generalized quasi-topological gravity}},  {\em JHEP} {\bf 08}
  (2019) 068 [\href{http://arXiv.org/abs/1902.02005}{{\tt 1902.02005}}].

\bibitem{Mir:2019rik}
M.~Mir and R.~B. Mann, {\it {On generalized quasi-topological cubic-quartic
  gravity: thermodynamics and holography}},  {\em JHEP} {\bf 07} (2019) 012
  [\href{http://arXiv.org/abs/1902.10906}{{\tt 1902.10906}}].

\bibitem{Edelstein:2022xlb}
J.~D. Edelstein, N.~Grandi and A.~Rivadulla~S\'anchez, {\it {Holographic
  superconductivity in Einsteinian Cubic Gravity}},  {\em JHEP} {\bf 05} (2022)
  188 [\href{http://arXiv.org/abs/2202.05781}{{\tt 2202.05781}}].

\bibitem{Murcia:2023zok}
A.~J. Murcia and D.~Sorokin, {\it {Universal aspects of holographic quantum
  critical transport with self-duality}},  {\em Phys. Rev. D} {\bf 108} (2023),
  no.~4 L041901 [\href{http://arXiv.org/abs/2303.12866}{{\tt 2303.12866}}].

\bibitem{Cano:2020ezi}
P.~A. Cano and {\'A}.~Murcia, {\it {Resolution of Reissner-Nordstr\"om
  singularities by higher-derivative corrections}},  {\em Class. Quant. Grav.}
  {\bf 38} (2021), no.~7 075014 [\href{http://arXiv.org/abs/2006.15149}{{\tt
  2006.15149}}].

\bibitem{Cano:2020qhy}
P.~A. Cano and {\'A}.~Murcia, {\it {Electromagnetic Quasitopological
  Gravities}},  {\em JHEP} {\bf 10} (2020) 125
  [\href{http://arXiv.org/abs/2007.04331}{{\tt 2007.04331}}].

\bibitem{Bueno:2021krl}
P.~Bueno, P.~A. Cano, J.~Moreno and G.~van~der Velde, {\it {Regular black holes
  in three dimensions}},  {\em Phys. Rev. D} {\bf 104} (2021), no.~2 L021501
  [\href{http://arXiv.org/abs/2104.10172}{{\tt 2104.10172}}].

\bibitem{Bueno:2022ewf}
P.~Bueno, P.~A. Cano, J.~Moreno and G.~van~der Velde, {\it {Electromagnetic
  generalized quasitopological gravities in (2+1) dimensions}},  {\em Phys.
  Rev. D} {\bf 107} (2023), no.~6 064050
  [\href{http://arXiv.org/abs/2212.00637}{{\tt 2212.00637}}].

\bibitem{Bueno:2025jgc}
P.~Bueno, O.~Lasso~Andino, J.~Moreno and G.~van~der Velde, {\it {On regular
  charged black holes in three dimensions}},
  \href{http://arXiv.org/abs/2503.02930}{{\tt 2503.02930}}.

\bibitem{Gross:1986iv}
D.~J. Gross and E.~Witten, {\it {Superstring Modifications of Einstein's
  Equations}},  {\em Nucl. Phys.} {\bf B277} (1986) 1.
%%CITATION = NUPHA,B277,1;%%

\bibitem{Gross:1986mw}
D.~J. Gross and J.~H. Sloan, {\it {The Quartic Effective Action for the
  Heterotic String}},  {\em Nucl. Phys.} {\bf B291} (1987) 41--89.
%%CITATION = NUPHA,B291,41;%%

\bibitem{Metsaev:1987zx}
R.~R. Metsaev and A.~A. Tseytlin, {\it {Order alpha-prime (Two Loop)
  Equivalence of the String Equations of Motion and the Sigma Model Weyl
  Invariance Conditions: Dependence on the Dilaton and the Antisymmetric
  Tensor}},  {\em Nucl. Phys. B} {\bf 293} (1987) 385--419.

\bibitem{Bueno:2019ltp}
P.~Bueno, P.~A. Cano, J.~Moreno and {\'A}.~Murcia, {\it {All higher-curvature
  gravities as Generalized quasi-topological gravities}},  {\em JHEP} {\bf 11}
  (2019) 062 [\href{http://arXiv.org/abs/1906.00987}{{\tt 1906.00987}}].

\bibitem{Cisterna:2018tgx}
A.~Cisterna, N.~Grandi and J.~Oliva, {\it {On four-dimensional Einsteinian
  gravity, quasitopological gravity, cosmology and black holes}},  {\em Phys.
  Lett. B} {\bf 805} (2020) 135435 [\href{http://arXiv.org/abs/1811.06523}{{\tt
  1811.06523}}].

\bibitem{Arciniega:2018fxj}
G.~Arciniega, J.~D. Edelstein and L.~G. Jaime, {\it {Towards geometric
  inflation: the cubic case}},  {\em Phys. Lett. B} {\bf 802} (2020) 135272
  [\href{http://arXiv.org/abs/1810.08166}{{\tt 1810.08166}}].

\bibitem{Arciniega:2018tnn}
G.~Arciniega, P.~Bueno, P.~A. Cano, J.~D. Edelstein, R.~A. Hennigar and L.~G.
  Jaime, {\it {Geometric Inflation}},  {\em Phys. Lett. B} {\bf 802} (2020)
  135242 [\href{http://arXiv.org/abs/1812.11187}{{\tt 1812.11187}}].

\bibitem{Arciniega:2019oxa}
G.~Arciniega, P.~Bueno, P.~A. Cano, J.~D. Edelstein, R.~A. Hennigar and L.~G.
  Jaime, {\it {Cosmic inflation without inflaton}},  {\em Int. J. Mod. Phys. D}
  {\bf 28} (2019), no.~14 1944008.

\bibitem{Edelstein:2020lgv}
J.~D. Edelstein, R.~B. Mann, D.~V. Rodr\'\i{}guez and A.~Vilar~L\'opez, {\it
  {Small free field inflation in higher curvature gravity}},  {\em JHEP} {\bf
  01} (2021) 029 [\href{http://arXiv.org/abs/2007.07651}{{\tt 2007.07651}}].

\bibitem{Edelstein:2020nhg}
J.~D. Edelstein, D.~V\'azquez~Rodr\'\i{}guez and A.~Vilar~L\'opez, {\it
  {Aspects of Geometric Inflation}},  {\em JCAP} {\bf 12} (2020) 040
  [\href{http://arXiv.org/abs/2006.10007}{{\tt 2006.10007}}].

\bibitem{Hohm:2019jgu}
O.~Hohm and B.~Zwiebach, {\it {Duality invariant cosmology to all orders in
  $\alpha$'}},  {\em Phys. Rev. D} {\bf 100} (2019), no.~12 126011
  [\href{http://arXiv.org/abs/1905.06963}{{\tt 1905.06963}}].

\bibitem{Bueno:2024zsx}
P.~Bueno, P.~A. Cano, R.~A. Hennigar and A.~J. Murcia, {\it {Regular black
  holes from thin-shell collapse}},  {\em Phys. Rev. D} {\bf 111} (2025),
  no.~10 104009 [\href{http://arXiv.org/abs/2412.02740}{{\tt 2412.02740}}].

\bibitem{Bueno:2024eig}
P.~Bueno, P.~A. Cano, R.~A. Hennigar and A.~J. Murcia, {\it {Dynamical
  Formation of Regular Black Holes}},  {\em Phys. Rev. Lett.} {\bf 134} (2025),
  no.~18 181401 [\href{http://arXiv.org/abs/2412.02742}{{\tt 2412.02742}}].

\bibitem{Bueno:2025gjg}
P.~Bueno, P.~A. Cano, R.~A. Hennigar, A.~J. Murcia and A.~Vicente-Cano, {\it
  {Regular black holes from Oppenheimer-Snyder collapse}},
  \href{http://arXiv.org/abs/2505.09680}{{\tt 2505.09680}}.

\bibitem{Garraffo:2008hu}
C.~Garraffo and G.~Giribet, {\it {The Lovelock Black Holes}},  {\em Mod. Phys.
  Lett.} {\bf A23} (2008) 1801--1818
  [\href{http://arXiv.org/abs/0805.3575}{{\tt 0805.3575}}].
%%CITATION = ARXIV:0805.3575;%%

\bibitem{Bueno:2016ypa}
P.~Bueno, P.~A. Cano, V.~S. Min and M.~R. Visser, {\it {Aspects of general
  higher-order gravities}},  {\em Phys. Rev.} {\bf D95} (2017), no.~4 044010
  [\href{http://arXiv.org/abs/1610.08519}{{\tt 1610.08519}}].
%%CITATION = ARXIV:1610.08519;%%

\bibitem{Bueno:2022lhf}
P.~Bueno, P.~A. Cano, Q.~Llorens, J.~Moreno and G.~van~der Velde, {\it {Aspects
  of three-dimensional higher curvatures gravities}},  {\em Class. Quant.
  Grav.} {\bf 39} (2022), no.~12 125002
  [\href{http://arXiv.org/abs/2201.07266}{{\tt 2201.07266}}].

\bibitem{Bueno:2017sui}
P.~Bueno and P.~A. Cano, {\it {On black holes in higher-derivative gravities}},
   {\em Class. Quant. Grav.} {\bf 34} (2017), no.~17 175008
  [\href{http://arXiv.org/abs/1703.04625}{{\tt 1703.04625}}].
%%CITATION = ARXIV:1703.04625;%%

\bibitem{Hennigar:2017ego}
R.~A. Hennigar, D.~Kubizňák and R.~B. Mann, {\it {Generalized
  quasitopological gravity}},  {\em Phys. Rev.} {\bf D95} (2017), no.~10 104042
  [\href{http://arXiv.org/abs/1703.01631}{{\tt 1703.01631}}].
%%CITATION = ARXIV:1703.01631;%%

\bibitem{DiFilippo:2024mwm}
F.~Di~Filippo, I.~Kol\'a\v{r} and D.~Kubiznak, {\it {Inner-extremal regular
  black holes from pure gravity}},  {\em Phys. Rev. D} {\bf 111} (2025), no.~4
  L041505 [\href{http://arXiv.org/abs/2404.07058}{{\tt 2404.07058}}].

\bibitem{Frolov:2024hhe}
V.~P. Frolov, A.~Koek, J.~P. Soto and A.~Zelnikov, {\it {Regular black holes
  inspired by quasitopological gravity}},  {\em Phys. Rev. D} {\bf 111} (2025),
  no.~4 044034 [\href{http://arXiv.org/abs/2411.16050}{{\tt 2411.16050}}].

\bibitem{Fernandes:2025fnz}
P.~G.~S. Fernandes, {\it {Singularity resolution and inflation from an infinite
  tower of regularized curvature corrections}},
  \href{http://arXiv.org/abs/2504.07692}{{\tt 2504.07692}}.

\bibitem{Fernandes:2025eoc}
P.~G.~S. Fernandes, {\it {Regular BTZ black holes from an infinite tower of
  corrections}},  \href{http://arXiv.org/abs/2504.08565}{{\tt 2504.08565}}.

\bibitem{Camanho:2011rj}
X.~O. Camanho and J.~D. Edelstein, {\it {A Lovelock black hole bestiary}},
  {\em Class. Quant. Grav.} {\bf 30} (2013) 035009
  [\href{http://arXiv.org/abs/1103.3669}{{\tt 1103.3669}}].
%%CITATION = ARXIV:1103.3669;%%

\bibitem{Hayward:2005gi}
S.~A. Hayward, {\it {Formation and evaporation of regular black holes}},  {\em
  Phys. Rev. Lett.} {\bf 96} (2006) 031103
  [\href{http://arXiv.org/abs/gr-qc/0506126}{{\tt gr-qc/0506126}}].

\bibitem{Dymnikova:1992ux}
I.~Dymnikova, {\it {Vacuum nonsingular black hole}},  {\em Gen. Rel. Grav.}
  {\bf 24} (1992) 235--242.

\bibitem{bardeen1968non}
J.~Bardeen, {\it Non-singular general relativistic gravitational collapse},  in
  {\em Proceedings of the 5th International Conference on Gravitation and the
  Theory of Relativity}, p.~87, 1968.

\bibitem{Wald:1993nt}
R.~M. Wald, {\it {Black hole entropy is the Noether charge}},  {\em Phys. Rev.}
  {\bf D48} (1993) 3427--3431 [\href{http://arXiv.org/abs/gr-qc/9307038}{{\tt
  gr-qc/9307038}}].
%%CITATION = GR-QC/9307038;%%

\bibitem{Iyer:1994ys}
V.~Iyer and R.~M. Wald, {\it {Some properties of Noether charge and a proposal
  for dynamical black hole entropy}},  {\em Phys. Rev.} {\bf D50} (1994)
  846--864 [\href{http://arXiv.org/abs/gr-qc/9403028}{{\tt gr-qc/9403028}}].
%%CITATION = GR-QC/9403028;%%

\bibitem{Tangherlini:1963bw}
F.~R. Tangherlini, {\it {Schwarzschild field in n dimensions and the
  dimensionality of space problem}},  {\em Nuovo Cim.} {\bf 27} (1963)
  636--651.
%%CITATION = NUCIA,27,636;%%

\bibitem{Myung:2006qr}
Y.~S. Myung, Y.-W. Kim and Y.-J. Park, {\it {Black hole thermodynamics with
  generalized uncertainty principle}},  {\em Phys. Lett. B} {\bf 645} (2007)
  393--397 [\href{http://arXiv.org/abs/gr-qc/0609031}{{\tt gr-qc/0609031}}].

\bibitem{Tharanath:2014naa}
R.~Tharanath, J.~Suresh and V.~C. Kuriakose, {\it {Phase transitions and
  Geometrothermodynamics of Regular black holes}},  {\em Gen. Rel. Grav.} {\bf
  47} (2015), no.~4 46 [\href{http://arXiv.org/abs/1406.3916}{{\tt
  1406.3916}}].

\bibitem{Dymnikova:2018uyo}
I.~Dymnikova, {\it {Generic Features of Thermodynamics of Horizons in Regular
  Spherical Space-Times of the Kerr-Schild Class}},  {\em Universe} {\bf 4}
  (2018), no.~5 63.

\bibitem{Koshelev:2024wfk}
A.~S. Koshelev and A.~Tokareva, {\it {Nonperturbative quantum gravity denounces
  singular black holes}},  {\em Phys. Rev. D} {\bf 111} (2025), no.~8 086026
  [\href{http://arXiv.org/abs/2404.07925}{{\tt 2404.07925}}].

\bibitem{Koshelev:2024lyu}
A.~S. Koshelev, C.~Li and A.~Tokareva, {\it {Quasi-normal modes in
  non-perturbative quantum gravity}},
  \href{http://arXiv.org/abs/2412.02678}{{\tt 2412.02678}}.

\bibitem{Konoplya:2024hfg}
R.~A. Konoplya and A.~Zhidenko, {\it {Infinite tower of higher-curvature
  corrections: Quasinormal modes and late-time behavior of D-dimensional
  regular black holes}},  {\em Phys. Rev. D} {\bf 109} (2024), no.~10 104005
  [\href{http://arXiv.org/abs/2403.07848}{{\tt 2403.07848}}].

\bibitem{Konoplya:2024kih}
R.~A. Konoplya and A.~Zhidenko, {\it {Dymnikova black hole from an infinite
  tower of higher-curvature corrections}},  {\em Phys. Lett. B} {\bf 856}
  (2024) 138945 [\href{http://arXiv.org/abs/2404.09063}{{\tt 2404.09063}}].

\bibitem{Birmingham:2001pj}
D.~Birmingham, I.~Sachs and S.~N. Solodukhin, {\it {Conformal field theory
  interpretation of black hole quasinormal modes}},  {\em Phys. Rev. Lett.}
  {\bf 88} (2002) 151301 [\href{http://arXiv.org/abs/hep-th/0112055}{{\tt
  hep-th/0112055}}].

\bibitem{Chan:1996yk}
J.~S.~F. Chan and R.~B. Mann, {\it {Scalar wave falloff in asymptotically
  anti-de Sitter backgrounds}},  {\em Phys. Rev. D} {\bf 55} (1997) 7546--7562
  [\href{http://arXiv.org/abs/gr-qc/9612026}{{\tt gr-qc/9612026}}].

\bibitem{Brandhuber:1998bs}
A.~Brandhuber, N.~Itzhaki, J.~Sonnenschein and S.~Yankielowicz, {\it {Wilson
  loops in the large N limit at finite temperature}},  {\em Phys. Lett. B} {\bf
  434} (1998) 36--40 [\href{http://arXiv.org/abs/hep-th/9803137}{{\tt
  hep-th/9803137}}].

\bibitem{Arias:2009me}
R.~E. Arias and G.~A. Silva, {\it {Wilson loops stability in the gauge/string
  correspondence}},  {\em JHEP} {\bf 01} (2010) 023
  [\href{http://arXiv.org/abs/0911.0662}{{\tt 0911.0662}}].

\bibitem{Baron:2013cya}
W.~H. Baron and M.~Schvellinger, {\it {Quantum corrections to dynamical
  holographic thermalization: entanglement entropy and other non-local
  observables}},  {\em JHEP} {\bf 08} (2013) 035
  [\href{http://arXiv.org/abs/1305.2237}{{\tt 1305.2237}}].

\bibitem{Grisaru:1986px}
M.~T. Grisaru, A.~E.~M. van~de Ven and D.~Zanon, {\it {Four Loop beta Function
  for the N=1 and N=2 Supersymmetric Nonlinear Sigma Model in Two-Dimensions}},
   {\em Phys. Lett.} {\bf B173} (1986) 423--428.
%%CITATION = PHLTA,B173,423;%%

\bibitem{Howe:1983sra}
P.~S. Howe and P.~C. West, {\it {The Complete N=2, D=10 Supergravity}},  {\em
  Nucl. Phys.} {\bf B238} (1984) 181--220.
%%CITATION = NUPHA,B238,181;%%

\bibitem{Myers:2008yi}
R.~C. Myers, M.~F. Paulos and A.~Sinha, {\it {Quantum corrections to eta/s}},
  {\em Phys. Rev.} {\bf D79} (2009) 041901
  [\href{http://arXiv.org/abs/0806.2156}{{\tt 0806.2156}}].
%%CITATION = ARXIV:0806.2156;%%

\bibitem{Buchel:2008ae}
A.~Buchel, R.~C. Myers, M.~F. Paulos and A.~Sinha, {\it {Universal holographic
  hydrodynamics at finite coupling}},  {\em Phys. Lett.} {\bf B669} (2008)
  364--370 [\href{http://arXiv.org/abs/0808.1837}{{\tt 0808.1837}}].
%%CITATION = ARXIV:0808.1837;%%

\bibitem{Galante:2013wta}
D.~A. Galante and R.~C. Myers, {\it {Holographic Renyi entropies at finite
  coupling}},  {\em JHEP} {\bf 08} (2013) 063
  [\href{http://arXiv.org/abs/1305.7191}{{\tt 1305.7191}}].
%%CITATION = ARXIV:1305.7191;%%

\bibitem{Gubser:1998nz}
S.~S. Gubser, I.~R. Klebanov and A.~A. Tseytlin, {\it {Coupling constant
  dependence in the thermodynamics of N=4 supersymmetric Yang-Mills theory}},
  {\em Nucl. Phys.} {\bf B534} (1998) 202--222
  [\href{http://arXiv.org/abs/hep-th/9805156}{{\tt hep-th/9805156}}].
%%CITATION = HEP-TH/9805156;%%

\bibitem{Buchel:2004di}
A.~Buchel, J.~T. Liu and A.~O. Starinets, {\it {Coupling constant dependence of
  the shear viscosity in N=4 supersymmetric Yang-Mills theory}},  {\em Nucl.
  Phys.} {\bf B707} (2005) 56--68
  [\href{http://arXiv.org/abs/hep-th/0406264}{{\tt hep-th/0406264}}].
%%CITATION = HEP-TH/0406264;%%

\bibitem{Kastor:2009wy}
D.~Kastor, S.~Ray and J.~Traschen, {\it {Enthalpy and the Mechanics of AdS
  Black Holes}},  {\em Class. Quant. Grav.} {\bf 26} (2009) 195011
  [\href{http://arXiv.org/abs/0904.2765}{{\tt 0904.2765}}].

\bibitem{Dolan:2010ha}
B.~P. Dolan, {\it {The cosmological constant and the black hole equation of
  state}},  {\em Class. Quant. Grav.} {\bf 28} (2011) 125020
  [\href{http://arXiv.org/abs/1008.5023}{{\tt 1008.5023}}].

\bibitem{Kubiznak:2016qmn}
D.~Kubiznak, R.~B. Mann and M.~Teo, {\it {Black hole chemistry: thermodynamics
  with Lambda}},  {\em Class. Quant. Grav.} {\bf 34} (2017), no.~6 063001
  [\href{http://arXiv.org/abs/1608.06147}{{\tt 1608.06147}}].

\bibitem{Ahmed:2023snm}
M.~B. Ahmed, W.~Cong, D.~Kubiz\v{n}\'ak, R.~B. Mann and M.~R. Visser, {\it
  {Holographic Dual of Extended Black Hole Thermodynamics}},  {\em Phys. Rev.
  Lett.} {\bf 130} (2023), no.~18 181401
  [\href{http://arXiv.org/abs/2302.08163}{{\tt 2302.08163}}].

\bibitem{Ahmed:2023dnh}
M.~B. Ahmed, W.~Cong, D.~Kubiznak, R.~B. Mann and M.~R. Visser, {\it
  {Holographic CFT phase transitions and criticality for rotating AdS black
  holes}},  {\em JHEP} {\bf 08} (2023) 142
  [\href{http://arXiv.org/abs/2305.03161}{{\tt 2305.03161}}].

\bibitem{Hajian:2023bhq}
K.~Hajian and B.~Tekin, {\it {Coupling Constants as Conserved Charges in Black
  Hole Thermodynamics}},  {\em Phys. Rev. Lett.} {\bf 132} (2024), no.~19
  191401 [\href{http://arXiv.org/abs/2309.07634}{{\tt 2309.07634}}].

\bibitem{Kastor:2010gq}
D.~Kastor, S.~Ray and J.~Traschen, {\it {Smarr Formula and an Extended First
  Law for Lovelock Gravity}},  {\em Class. Quant. Grav.} {\bf 27} (2010) 235014
  [\href{http://arXiv.org/abs/1005.5053}{{\tt 1005.5053}}].

\bibitem{Cai:2013qga}
R.-G. Cai, L.-M. Cao, L.~Li and R.-Q. Yang, {\it {P-V criticality in the
  extended phase space of Gauss-Bonnet black holes in AdS space}},  {\em JHEP}
  {\bf 09} (2013) 005 [\href{http://arXiv.org/abs/1306.6233}{{\tt 1306.6233}}].

\bibitem{Henriquez-Baez:2022bfi}
C.~Henr\'\i{}quez-B\'aez, J.~Oliva, M.~Oyarzo and M.~I. Y.~n. Reyes, {\it {R2
  corrections to the black string instability and the boosted black string}},
  {\em Phys. Rev. D} {\bf 107} (2023), no.~4 044021
  [\href{http://arXiv.org/abs/2212.07296}{{\tt 2212.07296}}].

\bibitem{Hennigar:2015esa}
R.~A. Hennigar, W.~G. Brenna and R.~B. Mann, {\it {$P-v$ criticality in
  quasitopological gravity}},  {\em JHEP} {\bf 07} (2015) 077
  [\href{http://arXiv.org/abs/1505.05517}{{\tt 1505.05517}}].

\bibitem{Hennigar:2025ftm}
R.~A. Hennigar, D.~Kubiz\v{n}\'ak, S.~Murk and I.~Soranidis, {\it
  {Thermodynamics of Regular Black Holes in Anti-de Sitter Space}},
  \href{http://arXiv.org/abs/2505.11623}{{\tt 2505.11623}}.

\bibitem{Paulos:2008tn}
M.~F. Paulos, {\it {Higher derivative terms including the Ramond-Ramond
  five-form}},  {\em JHEP} {\bf 10} (2008) 047
  [\href{http://arXiv.org/abs/0804.0763}{{\tt 0804.0763}}].

\bibitem{Pawelczyk:1998pb}
J.~Pawelczyk and S.~Theisen, {\it {AdS(5) x S**5 black hole metric at
  O(alpha-prime**3)}},  {\em JHEP} {\bf 09} (1998) 010
  [\href{http://arXiv.org/abs/hep-th/9808126}{{\tt hep-th/9808126}}].

\bibitem{Breitenlohner:1982bm}
P.~Breitenlohner and D.~Z. Freedman, {\it {Positive Energy in anti-De Sitter
  Backgrounds and Gauged Extended Supergravity}},  {\em Phys. Lett.} {\bf B115}
  (1982) 197.
%%CITATION = PHLTA,B115,197;%%

\bibitem{Breitenlohner:1982jf}
P.~Breitenlohner and D.~Z. Freedman, {\it {Stability in Gauged Extended
  Supergravity}},  {\em Annals Phys.} {\bf 144} (1982) 249.
%%CITATION = APNYA,144,249;%%

\bibitem{Chernicoff:2020tvr}
M.~Chernicoff, E.~Garc\'\i{}a, G.~Giribet and E.~Rub\'\i{}n~de Celis, {\it
  {Thin-shell wormholes in AdS$_5$ and string dioptrics}},  {\em JHEP} {\bf 10}
  (2020) 019 [\href{http://arXiv.org/abs/2006.07428}{{\tt 2006.07428}}].

\bibitem{Chernicoff:2022slp}
M.~Chernicoff, G.~Giribet and E.~R. de~Celis, {\it {Extremal surfaces and
  thin-shell wormholes}},  {\em Phys. Rev. D} {\bf 106} (2022), no.~8 086012
  [\href{http://arXiv.org/abs/2207.00072}{{\tt 2207.00072}}].

\bibitem{Arias:2010xg}
R.~E. Arias, M.~Botta~Cantcheff and G.~A. Silva, {\it {Lorentzian AdS,
  Wormholes and Holography}},  {\em Phys. Rev. D} {\bf 83} (2011) 066015
  [\href{http://arXiv.org/abs/1012.4478}{{\tt 1012.4478}}].

\bibitem{Lu:2013awa}
H.~L\"u, J.~F. Vazquez-Poritz and Z.~Zhang, {\it {Strings on AdS Wormholes and
  Nonsingular Black Holes}},  {\em Class. Quant. Grav.} {\bf 32} (2015), no.~2
  025005 [\href{http://arXiv.org/abs/1309.2957}{{\tt 1309.2957}}].

\bibitem{Anabalon:2025uzl}
A.~Anabal\'on, M.~Chernicoff, G.~Giribet, J.~Oliva and M.~Reyes, {\it
  {Quark-Antiquark Potential as a Probe for Holographic Phase Transitions}},
  \href{http://arXiv.org/abs/2501.15533}{{\tt 2501.15533}}.

\bibitem{Anabalon:2024lgp}
A.~Anabalon and J.~Oliva, {\it {Plasma-Plasma Third Order Phase Transition from
  Type IIB Supergravity}},  {\em Phys. Rev. Lett.} {\bf 133} (2024), no.~12
  121601 [\href{http://arXiv.org/abs/2405.04611}{{\tt 2405.04611}}].

\bibitem{Cisterna:2025vxk}
A.~Cisterna, M.~Hassaine and U.~Hernandez-Vera, {\it {Thermodynamics of
  four-dimensional regular black holes with an infinite tower of regularized
  curvature corrections}},  \href{http://arXiv.org/abs/2505.23467}{{\tt
  2505.23467}}.

\bibitem{Fierro:2020wps}
O.~Fierro, N.~Mora and J.~Oliva, {\it {Slowly rotating black holes in
  quasitopological gravity}},  {\em Phys. Rev. D} {\bf 103} (2021), no.~6
  064004 [\href{http://arXiv.org/abs/2012.06618}{{\tt 2012.06618}}].

\end{thebibliography}\endgroup

\end{document}